\documentclass[letterpaper,two column,longbibliography,notitlepage,superscriptaddress, accepted=2023-01-30]{quantumarticle}

\pdfoutput=1

\usepackage{float}
\usepackage{bbm}
\usepackage{epsfig}
\usepackage{epstopdf}
\usepackage{graphicx, subfigure}
\usepackage{amsmath,amssymb}
\usepackage[numbers,square]{natbib}
\usepackage{color}
\usepackage[colorlinks=true,citecolor=blue,linkcolor=magenta]{hyperref}
\usepackage[capitalize]{cleveref}
\usepackage{braket,mleftright}
\usepackage{hyperref}
\usepackage{dsfont}
\usepackage{extarrows}
\usepackage{physics}
\usepackage{nicefrac}
\usepackage{comment}
\usepackage{mathrsfs}

\DeclareMathAlphabet{\mathsfit}{T1}{\sfdefault}{\mddefault}{\sldefault}

\newcommand{\ie}{\textit{i.e.,~}}
\newcommand{\eg}{\textit{e.g.,~}}
\newcommand{\pp}{\mathsf{p}}
\newcommand{\ppbold}{\boldsymbol{\mathsf{p}}}
\newcommand{\ds}{\mathsf{d}}

\newcommand{\CC}{\mathsfit{C}}

\newcommand{\LL}{\mathsfit{L}}

\newcommand{\xx}{\mathsf{x}}
\newcommand{\xxbold}{\boldsymbol{\mathsf{x}}}

\newcommand{\onlinecite}[1]{\citenum{#1}}

\newcommand{\rev}[1]{{\color{black}#1}}

\begin{document}

\title{Quantum circuit debugging and sensitivity analysis via local inversions}
\author{Fernando A. Calderon-Vargas}
\email{facalde@sandia.gov}
\affiliation{Sandia National Laboratories, Livermore, CA 94550, USA}
\author{Timothy Proctor}
\affiliation{Quantum Performance Laboratory, Sandia National Laboratories,
Albuquerque, NM 87185, USA and Livermore, CA 94550, USA}
\author{Kenneth Rudinger}
\affiliation{Quantum Performance Laboratory, Sandia National Laboratories,
Albuquerque, NM 87185, USA and Livermore, CA 94550, USA}
\author{Mohan Sarovar}
\email{mnsarov@sandia.gov}
\affiliation{Sandia National Laboratories, Livermore, CA 94550, USA}

\begin{abstract}
As the width and depth of quantum circuits implemented by state-of-the-art quantum processors rapidly increase, circuit analysis and assessment via classical simulation are becoming unfeasible. \rev{It is crucial, therefore,} to develop new methods to identify significant error sources in large and complex quantum circuits. In this work, we present a technique that \rev{pinpoints the sections of a quantum} circuit that affect the circuit output the most and thus helps to identify the most significant sources of error. The technique requires no classical verification of the circuit output and is thus a scalable tool for debugging large quantum programs in the form of circuits. We demonstrate the practicality and efficacy of the proposed technique by applying it to example algorithmic circuits implemented on IBM quantum machines.
\end{abstract}

\maketitle

\section{Introduction}
Quantum computing hardware is undergoing rapid maturation and scaling, especially with superconducting circuit, trapped ion and, neutral atom platforms~\cite{IBM_127,Google_72,ball_physicists_2020,Rigetti}. In tandem, the development of increasingly sophisticated compilation tools makes it possible to write quantum programs in relatively high-level languages and execute them on quantum information processors (QIPs) with 20-50 qubits~\cite{aleksandrowicz_qiskit_2019,cirq_developers_cirq_2021,karalekas_pyquil_2020,svore_q_2018}. Successful execution, however, is not guaranteed due to the significant error rates present in today's, and likely near-future, QIPs~\cite{Preskill_2018}. Accurately predicting a program's success rate is increasingly difficult as the number of qubits and complexity of quantum circuits increases. This difficulty is because (i) classical simulation of ideal quantum circuits scales exponentially with qubit number and circuit depth, and (ii) predictions of circuit success from low-level characterization data, like individual gate errors, require unfeasible simulations of noisy quantum circuits or unreliable approximations.

This situation begs for developing efficient \emph{in-situ} debugging tools that enable reasoning about the source of errors in the output of a quantum circuit. How can one diagnose the most likely source of errors? Especially for QIPs with 50 or more qubits? This debugging task is made difficult by the fact that, unlike classical computers, the state of a QIP cannot be queried mid-execution without exponential resources (state tomography). 
Entirely new methods for debugging are therefore necessary to tackle this problem.

This work introduces a method for debugging quantum circuit implementations through a sensitivity analysis of the circuit output. Our method enables identifying which circuit layers most influence the circuit output and discovering time-dependent error modes such as degradation of gates as the circuit progresses. The method is based on locally inverting individual layers of a circuit.
Specifically, to study the errors induced by a particular ``target'' layer, we add two extra layers after it --a quasi-inverse of the target layer and the target layer again. In the absence of noise, these extra layers cancel each other and the circuit output is unaffected. In the presence of noise, however, the noise of the target layer is amplified. 
This local inversion is repeated for each circuit layer to be studied, and the circuit's output distribution is estimated for each version. 
By comparing these distributions, we can identify which layers perturb the circuit output the most. 

The practical utility of identifying the layers that influence the circuit output the most is \rev{the certainty that any effort to improve the gates in those layers will have the largest impact on the circuit outcome. For example, many of the quantum control techniques used to improve gate fidelity, like dynamically corrected gates~\cite{Khodjasteh2010,barnes_dynamically_2022} and composite pulse sequences~\cite{Merrill2012,Wang2012,Hill2007,Calderon-Vargas2016}, usually require longer gate times or long gate sequences, making them impractical to use in every noisy gate in a circuit. Knowing, however, the location of the most problematic gates in a circuit would allow us to efficiently use the quantum control techniques mentioned above.}

\rev{Surprisingly, the noisy gates affecting the circuit output the most} are not always the noisiest as identified by benchmarking of individual gates~\cite{knill2008randomized, magesan2011scalable, sheldon_characterizing_2016}.
We show that the impact of noise in a given layer is not only a function of the error rates of the individual gates that compose that layer but also of the algorithmic structure imposed by the entire circuit. Thus, the circuit-specific information we obtain about the impact of noise at a given layer on the circuit output cannot be reliably obtained from low-level characterization of individual gates, even in the absence of crosstalk or temporal instability. 
Our method is efficient and scalable in the low-noise regime; it requires the execution of circuits only slightly ($O(1)$) deeper than the original circuit and requires no classical computation of circuit outputs or simulation of noise models. 
Our technique does not require knowledge of whether the output is correct or not, nor does it provides information on its correctness.
The resource requirements and sample complexity of our debugging technique are discussed in \cref{sec:concl}.

Our technique should not be confused with methods for finding bugs in quantum programs~\cite{huang_statistical_2019,liu_quantum_2020,li_projection-based_2020,liu_systematic_2021} that use different assertions to help locate faulty statements in high-level specifications of quantum programs. Our method of local inversions, however, could be used in conjunction with quantum program assertions to readily locate bugs at multiple levels of quantum program specifications. We also note that the technique of local inversions of layers or individual gates has been proposed for error mitigation, in which case it is often referred to as \emph{unitary folding} \cite{Giurgica-Tiron_2020} or identity insertion \cite{McCaskey_2019, he_zero-noise_2020}. However, for error mitigation, the noise amplification provided by layer inversion is used for observable extrapolation and not for circuit debugging, as discussed in this work. 

We demonstrate our method using simulations and experiments, using circuits implementing the quantum approximate optimization algorithm (QAOA) for Max-Cut~\cite{Farhi2014} and quantum Fourier transform (QFT)~\cite{Nielsen2010}. This paper is organized as follows. In Sec.~\ref{sec:inv_intro}, we introduce the concept of local inversion and our analysis techniques. Section~\ref{sec:num_sim} presents numerical simulations demonstrating our debugging technique. In Sec.~\ref{sec:expts}, we present the results of the experimental implementation of our technique on cloud-accessible IBM QIPs. We give concluding remarks in Sec.~\ref{sec:concl}.

\section{Local inversions of a quantum circuit}
\label{sec:inv_intro}
Consider a quantum circuit $\CC$ represented as a series of $d$ layers, where each layer $\LL$ contains one or more quantum gates acting concurrently on an $n$-qubit system. 
We can write a $d$-layer circuit as 
\begin{equation}\label{eq:general_non_inverted_circuit}
\CC=\LL_{d} \LL_{d-1} \ldots \LL_{2}\LL_{1},
\end{equation}
Note that we only include gates in our definition of circuits and do not include state preparation and measurement (SPAM) quantum operations. Our strategy of amplifying the noise in circuit layers through local inversions cannot be applied to layers that include SPAM operations since these are non-invertible operations. As a result, our method cannot perform debugging or sensitivity analysis of SPAM operations. 

\begin{figure}[htb!]
\includegraphics[width=0.5\textwidth]{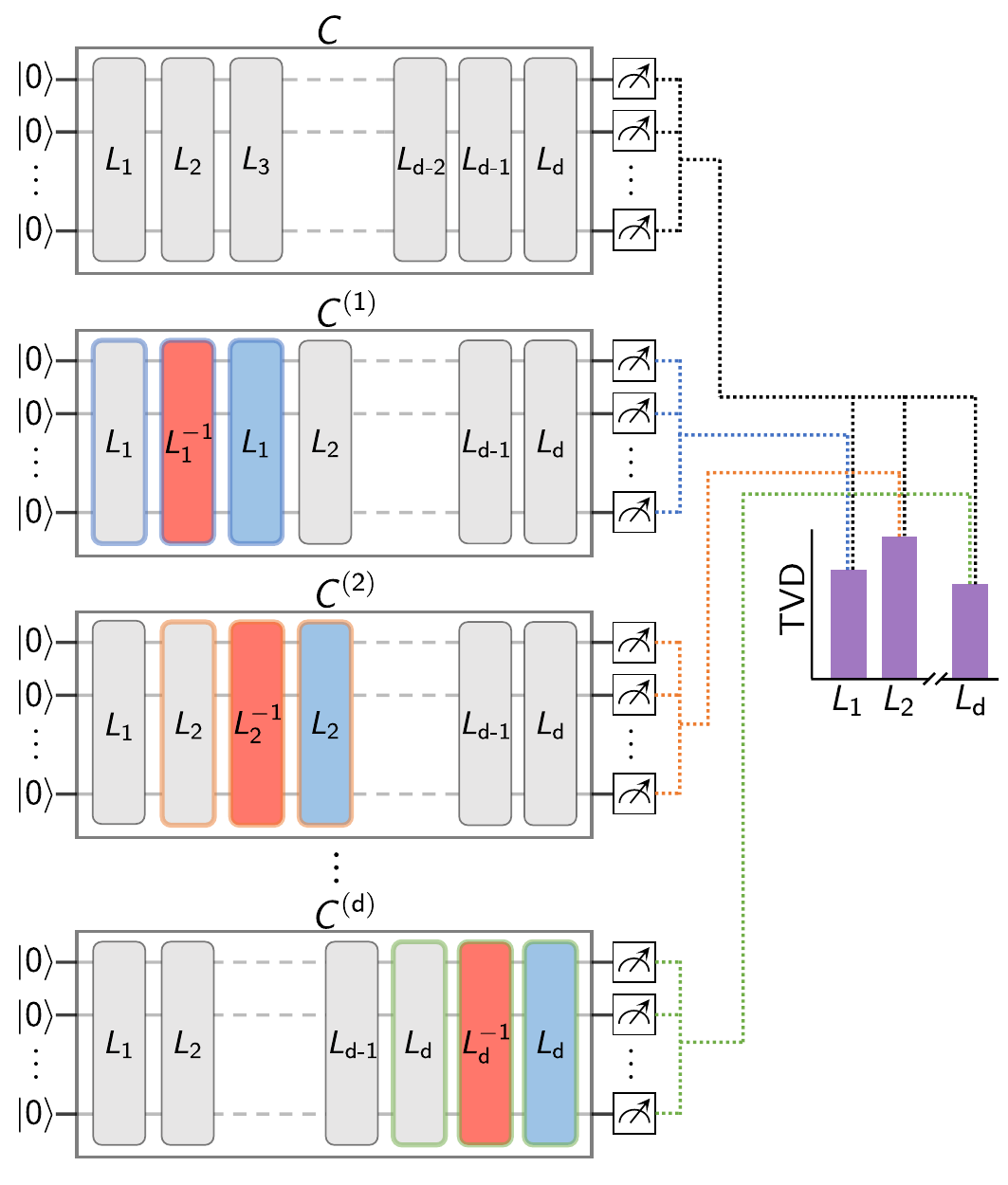}
\caption{The \textit{local inversion} technique shows the extent to which the errors in each of the $d$ layers ($\LL$) of a circuit $\CC$ perturb the output probability distribution. This is done by calculating the total variation distances (TVDs) between the output probability distribution of the circuit $\CC$ and the probability distributions of each of the $d$ circuits $\CC^{(i)}$. In each $\CC^{(i)}$ circuit, we invert the $i$-th layer and repeat it such that, in the absence of error on $\LL_i$, $\LL_i\LL_i^{-1}\LL_i=\LL_i$. The relative magnitude of the TVD between the probability distributions of $\CC$ and $\CC^{(i)}$ is proportional to the degree the circuit outcome is perturbed by the errors in layer $\LL_i$ (see \cref{app:Local_inversion_effect_on_prob_dist}).}
\label{fig:1}
\end{figure}

Each layer $\LL$ in Eq.~\eqref{eq:general_non_inverted_circuit}, in the absence of any noise, implements a unitary evolution $U(\LL)\in \text{SU}(2^n)$ on the $n$-qubit quantum state represented by the density operator $\rho$. Typically, some subset of noisy layers in a circuit $\CC$ affect the circuit outcomes the most. These dominant noisy layers are not necessarily those with the noisiest gates since the structure of the circuit can amplify or reduce the effect of errors on individual gates on the quantum state at the output of the circuit. We propose to use \textit{local inversions} to identify the dominant noisy layers (see Fig.~\ref{fig:1}). We  define the $i$-inverse of the circuit in Eq.~\eqref{eq:general_non_inverted_circuit} as
\begin{equation}\label{eq:general_inverted_circuit}
\begin{aligned}
\CC^{(i)}=&\LL_{d} \ldots \LL_{i+1}\left[\LL_{i}\LL^{-1}_{i}\LL_{i}\right] \LL_{i-1} \ldots \LL_{1},
\end{aligned}
\end{equation}
where we invert the $i$-th layer and repeat it. 
The inverse of a layer $i$, $\LL_i^{-1}$, is defined to be composed of gates that would ideally invert all the ideal gates in $\LL_i$. This means that, in the absence of error on the $i$-th layer, the outputs of both inverted and non-inverted circuits are the same. 
The actual composition of the inverse layers depends on the gate set available on the QIP. In some cases, $\LL^{-1}_i$ might be implemented as multiple physical layers because constructing the inverse of a native gate might require multiple native gates.
In the following, we work with the commonly used universal gate set $\{$\textsc{cnot}, $\sqrt{X}$, $Z(\theta)\}$, where the $Z(\theta)$ rotations are executed \textit{virtually} through a phase offset to subsequent drives, and thus take zero clock cycles and are almost error-free~\cite{McKay2017}. All layers are composed of some combination of these gates. The corresponding inverse layers, in this case, always consist of one physical layer since $\textsc{cnot}^{-1}=\textsc{cnot}$, $Z(\theta)^{-1}=Z(-\theta)$ and $\sqrt{X}^{-1} = Z(\pi)\sqrt{X}Z(-\pi)$. This final inverse requires only one physical layer since the $Z(\theta)$ gates are virtual.
In such cases, the inverse layer will have a similar error map as the original layer, and most likely, the error maps will be close \emph{in magnitude}.

We view $i$-inversions as local perturbations of the circuit and calculate how those local perturbations affect the circuit outcomes. To do so, we first define $\pp(k|\CC)$ as the probability distribution over measurement outcomes at the output of the circuit $\CC$, where $k \in B^{n}$ with $B^n$ being the set of all length $n$ bitstrings. We assume that all qubits are measured in the computational basis at the end of the circuit and that the input state is some fixed, but not necessarily known, initial state $\rho_0$  of all $n$ 	qubits. We then assess the impact of perturbations induced by $i$-inversion by computing the distance between the original and perturbed output distributions, $\ds[ \pp(\cdot|\CC), \pp (\cdot | \CC^{(i)})]$. In this work, we use the total variation distance (TVD) for the distance measure $\ds$ \footnote{We use the term ``distance measure'' loosely here. We do not demand that it is technically a distance metric, \ie satisfy symmetry and triangle inequality requirements.}. There are, however, many other choices for $\ds$, \eg Kullback-Leibler divergence and Jensen-Shannon divergence, each with a slightly different operational or statistical interpretation. In~\cref{app:qfim}, we detail a distance measure intimately related to sensitivity analysis that provides fine-grained information in the low-noise regime. However, since this regime is not necessarily realistic with current and near-future devices, we opt to use the TVD. Explicitly, 
\begin{equation}\label{eq:TVD_definition}
	\eta^{(i)} \equiv \ds_{\rm TVD}[\pp(\cdot|\CC), \pp(\cdot|\CC^{(i)})] = \frac{1}{2}\sum_k |\pp(k|\CC) - \pp(k|\CC^{(i)})|,
\end{equation}
which is interpreted as a measure of distinguishability of $\pp(\cdot|\CC)$ and $\pp(\cdot|\CC^{(i)})$; \ie the TVD is the greatest probability that one can discriminate two distributions based on one draw.

If, instead of inverting the $i$-th layer, we replace the gates in that layer with their ideal (noiseless) version, then  $\pp(k|\CC^{(i,\text{ideal})})$ would be the probability distribution of the $i$-ideal circuit  $\CC^{(i,\text{ideal})}$. We can then use  Eq.~\eqref{eq:TVD_definition} to obtain an expression for the TVD between $\pp(k|\CC^{(i,\text{ideal})})$ and the probability distribution of the noisy circuit $\CC$, ($\pp(k|\CC)$), obtaining:
\begin{equation}
\label{eq: TVD_with_ideal_layer}
	\eta^{(i,\text{ideal})} = \frac{1}{2}\sum_k |\pp(k|\CC) - \pp(k|\CC^{(i,\text{ideal})})|.	
\end{equation}

In \cref{app:Local_inversion_effect_on_prob_dist}, we show that if the error maps for the target layer and its inverse are similar, one can expect that $\eta^{(i)}\approx 2\eta^{(i,\text{ideal})}$. This provides some operational meaning to the TVD quantities we calculate through local inversion. Decreasing the error in layer $i$ will move the circuit output closer to the ideal output (in terms of TVD between output distributions) by roughly $\nicefrac{\eta^{(i)}}{2}$.

\rev{In this work, we assume that the circuit output is measured in the computational basis and evaluate errors using the TVD measure between outputs in this basis. For some quantum circuits, \eg primitives like the quantum Fourier transform, it might be important also to understand errors at the circuit output in other bases, \eg phase errors. Our technique can be easily modified to focus on such errors or average over possible types of errors by modifying the basis of measurement by adding a local Pauli layer at the end of the circuit.}

\section{Numerical simulations of circuit debugging via local inversion}
\label{sec:num_sim}
\begin{figure*}[htb!]
\includegraphics[width=1\textwidth]{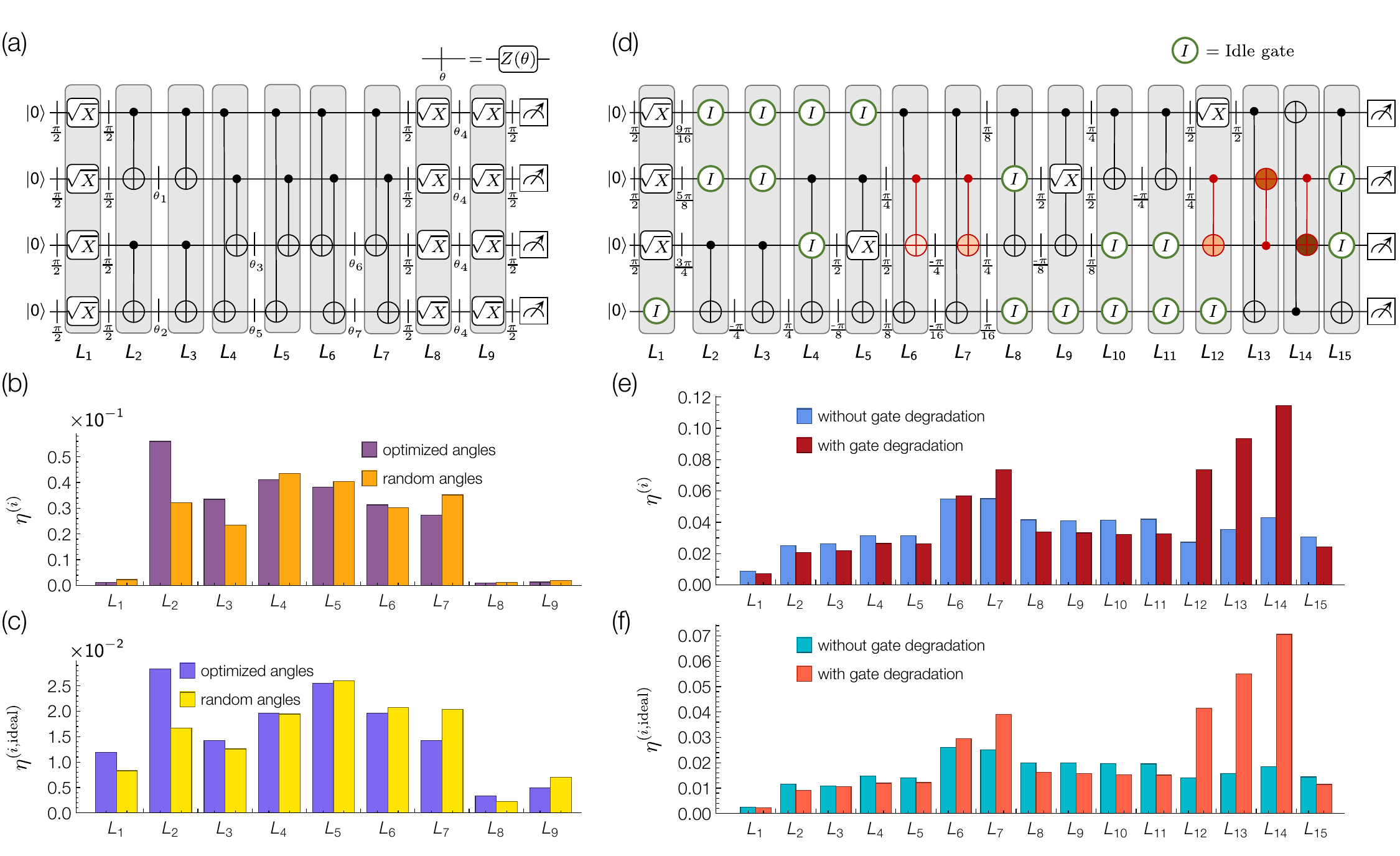}
\caption{\textbf{(a)} 4-qubit, one layer QAOA circuit for approximating solutions to Max-Cut on a weighted 4-vertex complete graph. The weights are $w_{(0,1)}=26$, $w_{(0,3)}=9$, $w_{(0,2)}=20$, $w_{(1,2)}=14$, $w_{(1,3)}=29$, $w_{(2,3)}=7$, where the subscripts indicate the respective graph edge. We consider two sets of $Z$-rotation angles (one is chosen randomly and used to find the second set via numerical optimization) to exemplify how changes in the angles of $Z$ rotations can influence which circuit layers affect the output distribution the most. The circuit is divided into nine layers. The values for the $Z$-rotation angles are given in Table~\ref{tab:set_angles_QAOA}. 
\textbf{(b)} TVD, Eq.~\eqref{eq:TVD_definition}, between the original and $i$-inverted circuit output distribution, obtained with the set of optimized (in violet) and random (in light orange) rotation angles. The layers with the largest TVD are those affecting the output distribution the most. Note that for the set of optimized angles the layer that is most influential is the second layer $\LL_2$, while for the set of random angles the fourth layer $\LL_4$ is the one affecting the output the most. 
\textbf{(c)} TVD  (Eq.~\eqref{eq: TVD_with_ideal_layer}) between the original and $i$-ideal circuit output distributions. The Pearson correlation coefficients (hereinafter correlations) between the TVDs, $\eta^{(i)}$ and $\eta^{(i,\text{ideal})}$, for the sets of optimized (in light purple) and random (in yellow) angles are 0.91 and 0.93, respectively. 
\textbf{(d)} 4-qubit QFT circuit used to show the layer inversion technique sensitivity to gate degradation during the circuit. We demonstrate this by adding depolarization channels to certain gates in the circuit. The circuit is divided into 15 layers. The first layer corresponds to the tensor product of Hadamard gates acting on the initial $\ket{0}^{\otimes 4}$ state to prepare an input state \rev{that gives a circuit output distribution that is peaked/concentrated in the computational basis, since the ideal circuit has a definite outcome}. The circuit also takes into account the idle gates present in each layer. The red shaded \textsc{cnot} gates include two-qubit depolarizing channels with increasing depolarizing probabilities; the larger the probability the darker the shade. 
\textbf{(e)} TVD,  Eq.~\eqref{eq:TVD_definition},  between the original and $i$-inverted circuit output distributions without (in blue) and with (in maroon) gate degradation.
\textbf{(f)}  TVD  (Eq.~\eqref{eq: TVD_with_ideal_layer}) between the original and $i$-ideal circuit output distributions. The correlations between the TVDs, $\eta^{(i)}$ and $\eta^{(i,\text{ideal})}$, for the circuits with  (in red) and without (in cerulean)  gate degradation are 0.997 and 0.99, respectively.}
\label{fig:2}
\end{figure*}

Before we present experimental results from the implementation of our circuit debugging procedure, we discuss some of the features of the approach using numerical simulations. We consider two quantum circuits: a 4-qubit circuit implementing the QAOA~\cite{Farhi2014} on a Max-Cut problem and a 4-qubit circuit implementing the QFT~\cite{Nielsen2010}. These circuits are shown in Figs.~\hyperref[fig:2]{\ref*{fig:2}(a)} and~\hyperref[fig:2]{\ref*{fig:2}(d)}, respectively. They are expressed in terms of the set of gates used by IBM's quantum processors: $\{ \textsc{cnot}, \sqrt{X}, Z(\theta)\}$. In these circuit diagrams, we denote virtual $Z(\theta)$ gates by a vertical line to indicate that they do not take up a clock cycle.
For the QAOA circuit, we have two sets of angles for the $Z$ rotations that define the QAOA variational angles, one random and the other optimized, both shown in Table~\ref{tab:set_angles_QAOA}. The parameters used for the optimized QAOA circuit were obtained with pyQAOA~\cite{pyQAOA}, a python package that optimizes QAOA circuits using gradient-based methods. 

\begin{table}[htb]
    \centering % instead of \begin{center}
    \caption{Sets of angles, labeled ``optimized'' and ``random'', used in the circuit implementing QAOA shown in Fig.~\hyperref[fig:2]{\ref*{fig:2}(a)}.}
    \vspace{1mm} % Adjust the height of the space between caption and tabular
\begin{tabular}{ |p{0.8cm}||p{1.3cm}|p{1.1cm}| }
 \hline
 Angles  & Optimized & Random  \\
 \hline
 $\theta_{1}$  & 4.426 &  4.6  \\
$\theta_{2}$ & 1.192 &  1.238 \\
$\theta_{3}$ & 2.383 &  2.477\\
$\theta_{4}$ & 3.8411 & 4.471\\
$\theta_{5}$ & 1.532 & 1.592  \\
$\theta_{6}$ & 3.404 & 3.538 \\
$\theta_{7}$ & 4.937 &  5.131  \\
 \hline
\end{tabular}\label{tab:set_angles_QAOA}
\end{table}

The noise model used in the simulations is Markovian and crosstalk-free, and is based on process matrices derived from performing one- and two-qubit gate-set tomography (GST) ~\cite{Blume-Kohout2017,pyGSTi} on the five-qubit quantum processor IBM-Q Ourense~\cite{cincio_machine_2021}. Each of the elementary gates, (\textsc{cnot}, $\sqrt{X}$, $I$), has a process matrix that has both coherent and stochastic errors, and these are explicitly presented in~\cref{app:error_model_Pauli_transfer_matrices}. We assume the $Z(\theta)$ gates are ideal. Although this noise model does not accurately capture all errors (\eg crosstalk), it is sufficient to demonstrate the important features of circuit debugging through local inversions.

The local inversion technique gauges the effect perturbations in each layer have on the circuit output distribution. The magnitude of such effects does not only depend on how noisy the gates present in that layer are, it also depends on which gates are present in the circuit, \ie the circuit structure, and for variational circuits, like in the QAOA case, the circuit parameters. 
Illustrating this, Fig.~\hyperref[fig:2]{\ref*{fig:2}(b)} shows the TVD, Eq.~\eqref{eq:TVD_definition}, between the original and $i$-inverse circuit output distribution for two almost identical 4-qubit QAOA circuits, the only difference being the values of the angles of their $Z$-rotations (see Table~\ref{tab:set_angles_QAOA}). Even though the difference in magnitude between the optimized and random angles is small, it is enough to modify which layers affect the output distribution the most. As shown in Fig.~\hyperref[fig:2]{\ref*{fig:2}(b)}, the second layer ($\LL_2$) is the dominant one when we use the set of optimized angles, while the dominant layer for the set of random angles is the fourth layer ($\LL_4$).

Next we demonstrate how local inversion technique can be used to detect gates with temporally varying noise (\ie drift). A common instance of this is gate degradation over the course of a circuit in some quantum computing platforms; \eg due to heating of vibrational modes M\o lmer-S\o reson gates can degrade in fidelity over the course of a long circuit. This is an effect that cannot be modeled using standard characterization methods that assume static process matrices. In Fig.~\hyperref[fig:2]{\ref*{fig:2}(e)} we demonstrate the signatures of gate degradation in $\eta^{(i)}$.  We simulate the 4-qubit circuit implementing a QFT, where in addition to the standard error model described above, we add two-qubit depolarizing channels to the \textsc{cnot} gates between qubits 1 and 2 (we label the first qubit  0), and increase the depolarizing probability with each successive application of the gate to model gate degradation (this increase is indicated by the darker shades of red in Fig.~\hyperref[fig:2]{\ref*{fig:2}(d)}). 
The depolarizing probability used for the red shaded \textsc{cnot} gates in layers 6, 7, 12, 13, and 14 are $\{p_1=0.025, p_2=0.051, p_3=0.076, p_4=0.102, p_5=0.127\}$, respectively. Figure~\hyperref[fig:2]{\ref*{fig:2}(e)} shows the TVDs for the circuits with and without gate degradation. The case where the \textsc{cnot} gate is degrading is clearly identified by an increasing TVD for later layers, thus showing the proposed technique's sensitivity to gate degradation during a circuit.

Finally, we comment on the correlation between $\eta^{(i)}$ and $\eta^{(i,\text{ideal})}$, where the latter is shown in Figs.~\hyperref[fig:2]{\ref*{fig:2}(c)} and~\hyperref[fig:2]{\ref*{fig:2}(f)}. These quantities should be correlated, especially in the weak-noise regime, and indeed the plots clearly show that $\eta^{(i)} \approx 2 \eta^{(i,\text{ideal})}$ for most $i$, for both example circuits. The notable deviation from this behavior is in layers 1, 8, and 9 of the QAOA circuit, where $\eta^{(i,\text{ideal})}$ is greater than $\eta^{(i)}$ (for both circuits with optimized and random angles). This mismatch in behavior is also reflected in measures of correlation between $\eta^{(i)}$ and $\eta^{(i,\text{ideal})}$: this correlation, for the QAOA circuit with optimized (random) angles is 0.91 (0.93), while for the QFT circuit with (without) depolarizing channels the correlation is 0.997 (0.99). 
To understand this mismatch, and thus the lower degree of correlation, note that layers 1, 8, and 9 of the QAOA circuit are composed entirely of $\sqrt{X}$ gates. As discussed in~\cref{app:error_model_Pauli_transfer_matrices}, the $\sqrt{X}$ gates have coherent noise that cancels when composed with its inverse. As a result, the layer inversion process does not amplify the noise in these layers as effectively as it does other layers. In contrast, the \textsc{cnot} gates are self-inverse and composing them does not result in any error cancellation with this error model. Below, we discuss a refinement of the basic layer inversion procedure presented above that mitigates such error cancellation and allows one to experimentally see the impact of noise in layers 1,8 and 9 on the circuit output. Meanwhile, note that since the QFT circuit has no layers composed entirely of $\sqrt{X}$ gates, and fewer $\sqrt{X}$ gates in general, such a mismatch between $\eta^{(i)}$ and $\eta^{(i,\text{ideal})}$ is not seen in Fig.~\hyperref[fig:2]{\ref*{fig:2}(f)}.

\subsection{Technique refinements: multiple layer inversion and averaging over random Pauli layers}\label{sec:technique_refinement}

The first refinement is the simple observation that one can perform multiple inversions of a layer and construct circuits of the form
\begin{equation}\label{eq:multiple_layer_inv_general_form}
\begin{aligned}
\CC^{(i)}=&\LL_{d} \ldots \LL_{i+1}\LL_{i}\left[\LL^{-1}_{i}\LL_{i}\right]^{m}\LL_{i-1} \ldots \LL_{1},
\end{aligned}
\end{equation}
where $m$ is the number of times the layer is being inverted. The effect of doing this is to amplify the perturbation from the target layer on the circuit output, which, as we demonstrate below, is useful for computing the sensitivity of the circuit output to the noise in a layer when it is small, or when there are sources of confounding noise like temporal instability of gate parameters on the timescale of the execution of a circuit debugging experiment. That is, consider that the total circuit debugging experiment consists of executing $d+1$ circuits, each $N_{\rm shots}$ times. Here, $d$ is the number of layers in the original circuit being debugged and $N_{\rm shots}$ is the number of measurements of each circuit. If gate parameters drift during this experiment time, then the observed difference between estimated $\pp(k|\CC)$ and $\pp(k|\CC^{(i)})$ could be due to this parameter drift rather than the amplified noise in layer $i$. 
In~\cref{app:drift} we discuss the techniques developed in Ref. \cite{rudinger_probing_2019} for detecting the presence of such temporal instability and post-selecting data to avoid confounding by drift.
In addition, performing multiple layer inversions, like described in \cref{eq:multiple_layer_inv_general_form} is a way of magnifying the effect of noise in layer $i$, which can be useful to increase the signal in the presence of small amounts of drift. 

\begin{figure}[htb!]
\includegraphics[width=0.46\textwidth]{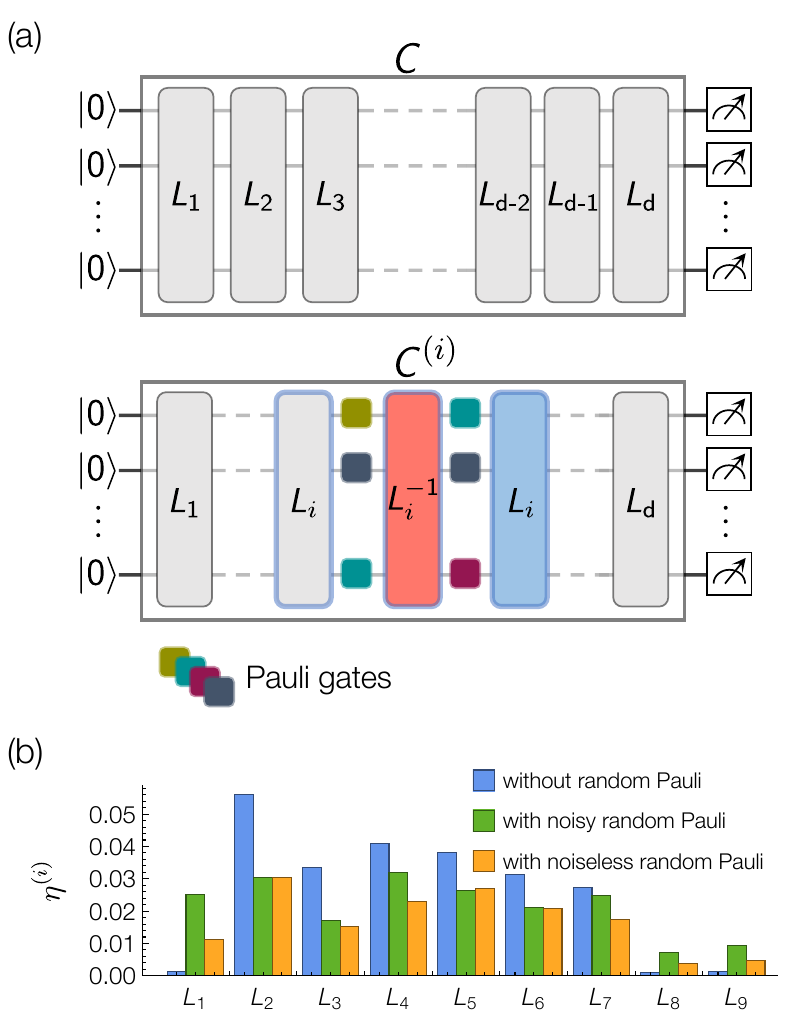}
\caption{\textbf{(a)} The introduction of randomly selected single-qubit Pauli gates before and after $\LL_i^{-1}$ (Pauli twirls) in an $i$-inverse circuit $\CC^{(i)}$ makes the TVD, between the output probability distribution of the original circuit $\CC$ and the average output distribution of $\CC^{(i)}$ over many Pauli twirls, sensitive to all coherent and incoherent noises.
\textbf{(b)} TVDs between the original and $i$-inverted 4-qubit QAOA circuit with optimized angles (Fig.~\hyperref[fig:2]{\ref*{fig:2}(a)}) $(i)$ without random Pauli gates (in blue), $(ii)$ with noisy random Pauli gates inserted during inversion (in green), and $(iii)$ with noiseless random Pauli gates inserted during inversion (in orange). In $(i)$ the basic local inversion technique fails to notice that there are noisy gates in the first and last two layers. With the addition of random Pauli gates during inversion (cases $(ii)$ and $(iii)$), our technique becomes sensitive again to the noise in the first and last two layers.}
\label{fig:3}
\end{figure}

To motivate the second refinement, note that the degree of noise amplification produced by the local inversion technique depends, in part, on the type of noise present in the target layer (see Eq.~\eqref{eq: Lambda operator}). For purely stochastic noise, the layer inversion readily amplifies the noise. Coherent noise, on the other hand, can lead to damping or cancellation~\cite{Hahn1950}, \eg when a gate over-rotates and its inverse under-rotates in the same amount they can cancel each other, leaving our technique insensitive to errors that do impact the circuit output distribution. 
To prevent this scenario, we refine our method by introducing a layer of random single-qubit Pauli gates before and after the inverse of each layer (see Fig.~\hyperref[fig:3]{\ref*{fig:3}(a)}) in such a way that, in the absence of noise, the circuit output is unaltered and, in the presence of coherent noise, the average over many such Pauli ``twirls'' removes any noise self-cancellation~\cite{proctor_measuring_2020}. Each time the circuit is executed, a new set of random Pauli gates are chosen such that the circuit takes the form
\begin{equation}\label{eq:multiple_layer_+_random_Pauli}
\begin{aligned}
\CC^{(i)}=&\LL_{d} \ldots \LL_{i} 
  \left[ \prod_{\ell=1}^m \mathcal{P}_{\ell}\LL^{-1}_{i}\tilde{\mathcal{P}}_{\ell} \LL_{i}\right] \ldots \LL_{1}, 
\end{aligned}
\end{equation}
where $m$ is the number of layer inversions, $\mathcal{P}_{\ell}$ is a layer of random single-qubit Pauli gates (chosen independently for each execution of the circuit), and $\mathcal{\tilde{P}}_{\ell}$ is another layer of single-qubit Pauli gates satisfying $\mathcal{P}_{\ell}\mathcal{U}^{\dagger}_{i}\mathcal{\tilde{P}}_{\ell} = \mathcal{U}^{\dagger}_{i}$ for all $\ell,i$. 
If local Pauli gates do not have negligible noise and are likely to introduce significant additional noise into the circuit, then it is preferable to minimize the number of random Paulis in between layer inversions. To achieve this, one can omit adding random single-qubit Pauli gates on qubits that are operated on by gates that have low chance of coherent error in $\LL_i$, \eg qubits that are idle or have a $Z(\theta)$ gate applied to them in layer $i$. We do this in the simulations and experiments described below.

To demonstrate this second refinement of the circuit debugging technique, we return to the QAOA circuit with optimized angles presented in Figs.~\hyperref[fig:2]{\ref*{fig:2}(a-c)}, where we find a mismatch between $\eta^{(i)}$ and $\eta^{(i,\text{ideal})}$ for layers composed entirely of $\sqrt{X}$ gates, and now include the random Pauli layers in the local inversions. 
We compare three cases: circuit debugging $(i)$ without random Pauli gates (same as in Fig.~\ref{fig:2}), $(ii)$ with noisy random Pauli gates (each Pauli is constructed using the noisy $\sqrt{X}$ gate and ideal $Z(\theta)$ rotations), and $(iii)$ with noiseless (perfect) random Pauli gates. For cases $(ii)$ and $(iii)$, we calculate the output probability distribution of 100 instances of the circuit, each with a random set of Pauli gates, and then take the average. Figure~\hyperref[fig:3]{\ref*{fig:3}(b)} clearly shows that, with random Pauli gates, either noisy or perfect, the local inversion technique correctly identifies the effect of the layers with coherent noise on the circuit output (layers 1,8 and 9). In fact, with the use of perfect random Pauli gates the correlation between $\eta^{(i)}$ and $\eta^{(i,\text{ideal})}$ (shown in Fig.~\hyperref[fig:2]{\ref*{fig:2}(c)}) increases from 0.91 to 0.99. The correlation when noisy random Pauli gates are used (0.84) is lower because the Pauli gates themselves introduce new noise that is not present in the original circuit. Therefore, we conclude that if single-qubit Pauli gates can be implemented with low noise, it is advantageous to perform the random Pauli twirling during layer inversions. 
 
Finally, we note that there is an alternative version of the first refinement that involves the insertion of $m$ repetitions of the target layer $\LL$, without any inversion, such that $\LL^m$ is equivalent to the identity in the absence of noise. This approach has the advantage that it removes the problem of  error cancellation. The only limitation of this approach is that a target layer containing different  types of gates will not necessarily be equal to the identity after $m$ repetitions. Of course, one can compile a circuit into a form where only one type of gate is applied in each layer, but this will increase the depth of circuits, which is usually not preferred, especially for near-term QIPs where it is desired to minimize circuit depth.

\section{Experimental demonstration of circuit debugging}\label{sec:expts}
\begin{figure*}[htb!]
\includegraphics[width=1\textwidth]{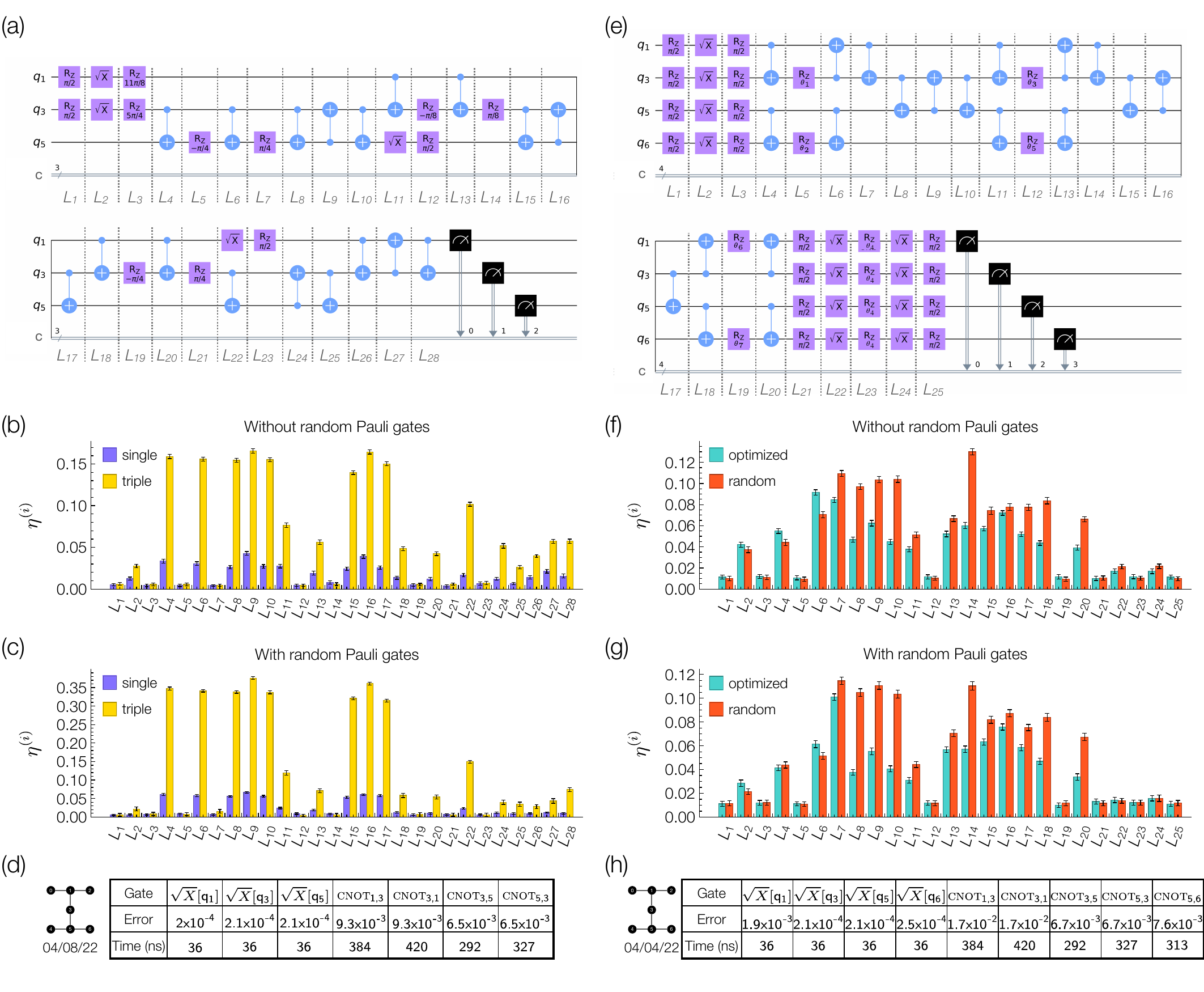}
\caption{\textbf{(a)} 3-qubit QFT circuit transpiled for the IBM QIP ibmq\_jakarta, active qubits \{1,3,5\}.
\textbf{(b)}  TVD for a single-layer inversion  (in light purple) and triple-layer inversion (in gold).
\textbf{(c)} TVD for a single-layer inversion  (in light purple) and triple-layer inversion (in gold) with random Pauli gates.
\textbf{(d)} Qubit layout and connectivity of ibmq\_jakarta, along with  the gate errors and gate times at the time of the QFT circuit debugging experiment.
\textbf{(e)} 4-qubit QAOA circuit transpiled for ibmq\_jakarta, active qubits \{1,3,5,6\}. 
\textbf{(f)} TVD for a triple layer inversion with the set of \textit{optimized} angles (in turquoise) and triple layer inversion with the set of \textit{random} angles (in red), see Table~\ref{tab:set_angles_QAOA}.
\textbf{(g)} TVD for a triple layer inversion  with the set of \textit{optimized} angles  (in turquoise) and triple layer inversion with the set of \textit{random} angles (in red), both with random Pauli gates. 
\textbf{(h)} Gate errors and gate times for ibmq\_jakarta at the time of the QAOA circuit debugging experiment.
The standard deviation for each of the TVDs, indicated with errors bars in (b), (c), (f), and (g), is calculated using non-parametric bootstrapping.
}
\label{fig:4}
\end{figure*}

The experimental results presented in this section were obtained using the IBM QIP ibmq\_jakarta. This QIP has the qubit layout and connectivity shown in Fig.~\hyperref[fig:4]{\ref*{fig:4}(d)}. Using qubits 1, 3, and 5 on ibmq\_jakarta we implemented a 3-qubit QFT circuit, and we used qubits 1, 3, 5, and 6 to implement a 4-qubit QAOA for Max-Cut circuit using two different sets of angles (see Table~\ref{tab:set_angles_QAOA}). The gate errors and gate times for IBM ibmq\_jakarta at the time of the QFT and QAOA experiments are presented in Figs.~\hyperref[fig:4]{\ref*{fig:4}(d)} and~\hyperref[fig:4]{\ref*{fig:4}(h)}, respectively.  In \cref{app:details_experiments} we detail the steps involved in these experiments.

Similarly to the QFT circuit in Fig.~\hyperref[fig:2]{\ref*{fig:2}(a)}, the first layer of the  3-qubit QFT circuit in Fig.~\hyperref[fig:4]{\ref*{fig:4}(a)} is a tensor product of Hadamard gates to prepare an input state \rev{that gives a circuit output distribution that is peaked/concentrated in the computational basis, since the ideal circuit has a definite outcome~\cite{QFT_state}}. For this circuit, we obtained experimental results using single and triple local inversion ($m=1$ and $m=3$, respectively) for circuits without  (Fig.~\hyperref[fig:4]{\ref*{fig:4}(b)}) and with random Pauli gates (Fig.~\hyperref[fig:4]{\ref*{fig:4}(c)}).
From the gate errors and gate times shown in the table in Fig.~\hyperref[fig:4]{\ref*{fig:4}(d)}, one would expect a layer like $\LL_{11}$, which has a $\sqrt{X}$ and the noisiest \textsc{cnot} in the circuit, to be among the layers with the largest TVD. Instead, in all cases, $\eta^{(11)}$ is dominated by $\eta^{(4)}, \eta^{(6)}, \eta^{(9)}, \eta^{(16)}$, which correspond to layers with a single \textsc{cnot} between qubits 3 and 5, which is not the noisiest nor the longest gate according to the table in Fig.~\hyperref[fig:4]{\ref*{fig:4}(d)}. In fact, for all cases, except the $m=1$, without random Paulis experiment (light purple in Fig.~\hyperref[fig:4]{\ref*{fig:4}(b)}), $\eta^{(11)}$ is also dominated by $\eta^{(8)},\eta^{(10)}, \eta^{(15)}, \eta^{(17)}$, which also correspond to layers with a single \textsc{cnot} between qubits 3 and 5. The increased sensitivity of the circuit output to this gate could be due to the circuit structure, or because the actual error rates for the \textsc{cnot} gates during these experiments were different from the reported values after calibration (\eg due to drift). This is an example of the useful information our circuit debugging technique provides, which cannot be inferred from low-level characterization information about each of the gates present in the circuit. 
In Figs.~\hyperref[fig:4]{\ref*{fig:4}(b)} and~\hyperref[fig:4]{\ref*{fig:4}(c)} we also see the effect of multiple local inversions: the TVDs from the triple local inversion experiments are considerably larger than those from the single local inversion one. \rev{This large difference in magnitude between the TVDs from triple and single local inversions is most likely due to non-negligible higher-order error terms of the error map (see Eq.~\ref{eq:varepsilon_G_i})}. Note also that, even though the triple inversion enlarges the TVD in each layer, the TVDs for layers with only $Z$-rotations are not enlarged. 
This is expected since the layers with only $Z(\theta)$ gates are not actually being locally inverted in the QIP (see \cref{app:details_experiments}), and any TVD change is likely due to temporal instability of the QIP. In \cref{app:qfim}, we discuss the effect of multiple local inversions using another distance measure and some of the advantages and disadvantages it presents. \rev{Lastly, Figure~\hyperref[fig:4]{\ref*{fig:4}(c)} shows the TVDs for a single-layer inversion and triple-layer inversion with random Pauli gates, of which many are twice as large as those without random Pauli gates shown in Fig.~\hyperref[fig:4]{\ref*{fig:4}(b)}. This large difference in magnitude seems to be a combination of coherent noise and new noise introduced by the random Pauli gates.}

Figures~\hyperref[fig:4]{\ref*{fig:4}(e-g)} present the experimental results for debugging of the 4-qubit QAOA circuit with both optimized and random angles (see Table~\ref{tab:set_angles_QAOA}). We show TVDs between the original and the $i$-inverted circuits without and with (Figs.~\hyperref[fig:4]{\ref*{fig:4}(f)} and~\hyperref[fig:4]{\ref*{fig:4}(g)}, respectively) random Pauli layers inserted during the inversion.  We show here only the results for triple layer inversion ($m=3$) for conciseness. 
The first observation is that depending on the QAOA angles used (optimized or random), the dominant layers can be different. This is similar to what was observed in the numerical simulation shown in Fig.~\hyperref[fig:2]{\ref*{fig:2}(b)}, where the output of two circuits with the same structure but slightly different set of $Z$-rotations can be sensitive to noise in different layers. This experimental result reaffirms that the circuit structure has a non-negligible influence on which layers end up influencing the circuit outcome the most. 
For the optimized angles, the three most dominant layers with and without the random Pauli gates are $\{\LL_{6}, \LL_{7}, \LL_{16}\}$ (closely followed by $\LL_{15}$ in the case with random Pauli gates), and for the random angles, the five most dominant layers are $\{\LL_7, \LL_8, \LL_9, \LL_{10}, \LL_{14}\}$.
Interestingly, only layers $\LL_6, \LL_7$, and $\LL_{14}$ have \textsc{cnot} gates between qubits 1 and 3, which are the noisiest and longest gates according to the calibration data table shown in Fig.~\hyperref[fig:4]{\ref*{fig:4}(h)}, the rest of the dominant layers  have \textsc{cnot} gates between qubits 3 and 5, which are the least noisy. Thus, although noise in the highest error gate affects the output distribution, other gates also have significant influence. Note as well that, with the exception of layer $\LL_6$, none of the layers with two \textsc{cnot} gates (the two noisiest gates according to the table in Fig.~\hyperref[fig:4]{\ref*{fig:4}(h)}) are among the layers with largest TVD. 
This is another example of the practical utility of the local inversion technique, it provides circuit-specific information about the impact of noisy gates on the output distribution that cannot be reliably obtained from low-level characterization of individual gates.

\section{Discussion}\label{sec:concl}
We have introduced a new method for debugging quantum circuit implementations based on \textit{local inversion} of individual layers of a circuit. Through analysis, simulation, and experiment, we have demonstrated that this technique enables the identification of circuit layers that affect the circuit output the most and detects gate degradation and temporal instability.

Moreover, we have shown that the impact of a given layer's error depends not only on the error rates of the gates that form the layer but also on the structure of the entire circuit. As such, the technique can be used to determine the source of errors in the output of quantum algorithmic circuits. 

\rev{We note that our technique can extract error behavior that is inefficient or impossible to derive from the characterization of individual gates, which is the dominant mode of quantum computer characterization at present. For example, the technique captures emergent error modes such as temporal variation of gates or dependence on gate behavior on neighboring gates. Such emergent error modes can be pathological, but if they are relevant to the algorithmic circuit that is being implemented, our technique will detect them. Moreover, our technique has the advantage of not needing classical simulation of circuits in order to extract these emergent error modes.}

The resources required for circuit debugging based on layer inversions generally scale favorably with circuit width ($n$, number of qubits) and depth ($d$). In addition to the execution of the original circuit to be debugged, $\CC$, the technique requires the execution of $d$ modified circuits, $\CC^{(i)}, 1\leq i \leq d$, of depth $d+\mathcal{O}(1)$. The number of executions of each circuit (measurement shots) is the only quantity with slightly complex scaling. Consider that underlying the debugging technique is the task of learning two distributions -- $\pp(\cdot|\CC)$ and $\pp(\cdot | \CC^{(i)})$ -- and computing the distance between them. The sample complexity of learning an arbitrary discrete distribution over $2^n$ elements to $\epsilon$ $\mathds{L}_1$ error is $\Theta(2^n/\epsilon^2)$, and the worst-case distributions to learn are typically those with high entropy. Therefore, the number of circuit executions required for our circuit debugging protocol scales exponentially in the worst case, which physically will correspond to cases where either $(i)$ the algorithmic circuit $\CC$ produces a close-to-uniform distribution over outcomes or $(ii)$ the circuit implementations are so noisy that the output is uniformly distributed across the outcomes (\eg a completely depolarized output state). We argue that in most cases of interest, the output distribution will have a low entropy because $(i)$ quantum algorithms typically produce low-entropy probability distributions over the measurement bit strings, and $(ii)$ any algorithmic circuit for which one would perform debugging would be in the low-noise regime, where the output is not completely depolarized. In such cases, the sample complexity of distribution learning is much more favorable. In fact, for a distribution over $2^n$ elements with $k$ \emph{modes} (loosely, peaks or valleys), learning to $\epsilon$ $\mathds{L}_1$ error requires poly($k$,$\log(n),1/\epsilon)$ samples \cite{daskalakis_learning_2014}. Moreover, there is also the alternative of combining the local inversion with mirror circuits, which can be used to obtain low-entropy output probability distributions~\cite{proctor_measuring_2020}. Therefore, in cases of practical application, none of the resources required for circuit debugging via local inversions scale super-polynomially with $n$ or $d$. 

We note that our method could be combined with error-mitigation techniques like the zero-noise extrapolation~\cite{temme_error_2017,kandala_error_2019,he_zero-noise_2020} \rev{and probabilistic error cancellation~\cite{temme_error_2017,van2022probabilistic}}. For zero-noise extrapolation, local inversion could be used to identify the layer/gate that perturbs the output distribution the most and use that layer/gate as a starting point for noise amplification and extrapolation. \rev{For probabilistic error cancellation, the main limitation is the sampling overhead, which depends on the number of qubits as well as the number of noisy layers. By focusing only on the layers affecting the outcome the most, our technique could potentially lead to lower sampling overheads without undermining the error mitigation. Finally, our technique can be used to assess the relative merits of different compilations of a quantum algorithm -- \ie hardware- and noise-aware compilations attempt to reduce the effect of hardware noise or limited connectivity on quantum algorithm accuracy and our debugging technique can be used to assess their effectiveness.}

\rev{The deployment of the circuit debugging tool we have developed to variational quantum circuits bears commenting upon. As we have shown, the most influential layer for a circuit depends on the values of the circuit parameters. That influential layer can, therefore, change depending on where one is in the variational parameter landscape during an optimization. Thus, the debugging protocol should be applied at each parameter value encountered during the optimization. We note that this is consistent with how error mitigation is typically performed for variational circuits -- the variational cost function is error mitigated at each value of the variational parameters. Therefore, as mentioned above, one could use our protocol at each step in the parameter optimization to identify the layers or gates to amplify during error mitigation.}

\rev{Finally, we note that Patel \emph{et al.} \cite{patel_2022} have also recently developed a similar protocol for ``identifying the most critical operations in quantum circuits'' through local inversions. We note that several elements distinguish our work from Ref. \cite{patel_2022}: (i) we present theoretical analysis to justify the local inversion technique and present an operationally meaningful interpretation of the experimentally measured quantity, (ii) we develop the Pauli twirling refinement to mitigate error cancellation effects that might lead to erroneous conclusions, and (iii) we incorporate drift detection into the debugging protocol in order to minimize conflation by temporal variation of gates.}

We conclude by noting that as QIPs mature and naturally attempt to execute increasingly complex quantum algorithms, lightweight, \emph{in-situ} diagnostics and debugging tools such as the one we introduced in this work will be essential. 

\section*{Acknowledgements}
This work was supported by the U.S.~Department of Energy (DOE) Office of Science Advanced Scientific Computing Research program office Accelerated Research for Quantum Computing (ARQC) program. 
Sandia National Laboratories is a multimission laboratory managed and operated by National Technology and Engineering Solutions of Sandia, LLC, a wholly owned subsidiary of Honeywell International, Inc., for the U.S. Department of Energy's National Nuclear Security Administration under contract DE-NA0003525. This paper describes objective technical results and analysis. Any subjective views or opinions that might be expressed in the paper do not necessarily represent the views of the U.S. Department of Energy or the United States Government.
We acknowledge the use of IBM Quantum services for this work. The views expressed are those of the authors, and do not reflect the official policy or position of IBM or the IBM Quantum team.

\nocite{apsrev42Control}
\bibliography{references}

\begin{thebibliography}{49}
\bibitem [{QFT()}]{QFT_state}%
  \BibitemOpen
  \href@noop {} {}\bibinfo {note} {When the input state is (or close to) a
  computational basis state, the output distribution given by QFT is close to
  uniform, rendering our technique inefficient as discussed in the Discussion
  section.}\BibitemShut {Stop}%
\bibitem [{vec()}]{vectorization_basis}%
  \BibitemOpen
  \href@noop {} {}\bibinfo {note} {The vectorization of a $k\times \ell$ matrix
  $A$ can be represented as a linear sum $\mathrm{vec}(A) = \sum_{i=1}^{\ell}
  \mathbf{e}_i \otimes A \mathbf{e}_i$, where $\mathbf{e}_i = [0,0,\ldots
  ,1,\dots ,0]^{\mathrm{T}}]$ is the $i$-th canonical basis for the
  $\ell$-dimensional space. The vectorized form of the matrix multiplication is
  given by $\mathrm{vec}(ABC)=(C^{\mathrm{T}}\otimes A)\mathrm{vec}(B)$, where
  $B$ ($C$) is an $\ell \times m$ ($m\times n$) matrix.}\BibitemShut {Stop}%
\end{thebibliography}

\clearpage

\begin{appendix}
\section{Analysis of local inversions in the low-error regime}
\label{app:Local_inversion_effect_on_prob_dist}
It is instructive to examine the behavior of $\eta^{(i)}$ (Eq.~\eqref{eq:TVD_definition}) perturbatively, in the low-error regime.
In order to do so, we find it convenient to use the vectorized form of $\rho$, \ie $\ket{\rho}\rangle:=\text{vec}(\rho)=(\rho_{11},\rho_{21},\ldots,\rho_{n1},\rho_{12},\ldots,\rho_{nn})^T$, then $\text{vec}(U\rho U^{\dagger})=(\bar{U}\otimes U)\text{vec}(\rho) = \mathcal{U}\ket{\rho}\rangle$~\cite{vectorization_basis}, where $\mathcal{U}=\bar{U}\otimes U$ and $\bar{U}$ is the complex conjugate of $U$. Note that, when noise is present, each layer $\LL$ actually implements $(\mathcal{E}\circ\mathcal{U})\ket{\rho}\rangle$, where $\mathcal{E}$ is a completely positive trace preserving \emph{error map}. Then, the probability distribution for a depth-$d$ circuit $\CC$, for ${k\in B^n}$, is given by
\begin{equation}\label{eq:p(kC)_basic}
\begin{aligned}
\pp(k|\CC) =& \langle\langle k\vert \CC \vert \rho_0\rangle\rangle\\
=&\langle\langle k\vert \mathcal{E}_{d}\mathcal{U}_{d}\cdots\mathcal{E}_{1}\mathcal{U}_{1}\vert \rho_0\rangle\rangle,
\end{aligned}
\end{equation}
where $\rho_0$ is the initial state of all $n$ qubits and $\langle\langle k\vert$ is the vectorized (non-ideal) POVM element that corresponds to measuring bit string $k$.
We can pull all the error maps to the left, obtaining
\begin{equation}\label{eq:prob_distr_E'_d...E'_1.C_0}
\pp(k|\CC)=\langle\langle k\vert  \mathcal{E}'_{d}\cdots\mathcal{E}'_{1}\CC_0\vert \rho_0\rangle\rangle,
\end{equation}
where $\CC_0=\mathcal{U}_{d}\cdots\mathcal{U}_{1}$ is the error-free circuit and 
\begin{equation}
\mathcal{E}'_{i} =( \mathcal{U}_{d}\cdots \mathcal{U}_{i+1}) \mathcal{E}_{i} ( \mathcal{U}_{i+1}^{\dagger}\cdots \mathcal{U}_{d}^{\dagger}).  
\end{equation}
We assume that the error map $\mathcal{E}_{i}$ in each layer is always close to the identity process $\mathds{1}$. The error map in the $i$-th layer can be written as~\cite{blume-kohout_taxonomy_2022}
\begin{equation}
\mathcal{E}_{i} = \exp(\mathscr{E}_i),
\end{equation}
where $\mathscr{E}_i$ is the error map's \textit{error generator}. Given that the error map is close to the identity, we can expand it as follows:
\begin{equation}\label{eq:varepsilon_G_i}
\mathcal{E}_{i}  \approx \mathds{1} + \mathscr{E}_i + \mathcal{O}(d^2\delta^2),
\end{equation}
where $\delta\equiv \vert\vert \mathscr{E}_i\vert\vert_{\diamond}$ \rev{and $d$ is the circuit's depth}. Then, to first order in $\delta$, Eq.~\eqref{eq:prob_distr_E'_d...E'_1.C_0} becomes:
\begin{equation}\label{eq:prob_distr_without_inversion}
\pp(k|\CC) \approx \pp(k|\CC_0) + \sum_i  \langle\langle k\vert\mathscr{E}_i^{'} \CC_0\vert\rho\rangle\rangle + \mathcal{O}(d^2\delta^2),
\end{equation}
where $\mathscr{E}_i^{'} =( \mathcal{U}_{d}\cdots \mathcal{U}_{i+1}) \mathscr{E}_i ( \mathcal{U}_{i+1}^{\dagger}\cdots \mathcal{U}_{d}^{\dagger})$.

We can follow the same analysis for the probability distribution over outcomes for the circuit $\CC^{(i)}$ (Eq.~\eqref{eq:general_inverted_circuit}) with the $i$-th layer inverted, and obtain the following
\begin{equation}\label{eq: prob_distri_local_inversion_Delta}
\begin{aligned}
\pp(k|\CC^{(i)}) & \approx \pp(k|\CC_0) + \sum_i  \langle\langle k\vert\mathscr{E}_i^{'}\CC_0\vert\rho\rangle\rangle + \Delta_k^{(i)} + \mathcal{O}(d^2\delta^2)\\
& \approx \pp(k|\CC) + \Delta_k^{(i)} +\mathcal{O}(d^2\delta^2),
\end{aligned}
\end{equation}
where
\begin{equation}\label{eq: Delta_inversion_error}
\begin{aligned}
\Delta_k^{(i)} = \langle\bra{k}&( \mathcal{U}_{d}\cdots \mathcal{U}_{i+1}) (\mathscr{E}_i + \mathcal{U}_{i}\mathscr{E}_i^{(-1)}\mathcal{U}_{i}^{\dagger})\\
& ( \mathcal{U}_{i+1}^{\dagger}\cdots \mathcal{U}_{d}^{\dagger}) \CC_0\ket{\rho_0}\rangle.
\end{aligned}
\end{equation}
Here $\mathcal{E}_{i}^{(-1)}$ is the error map for the layer that inverts layer $i$ and $\mathscr{E}_i^{(-1)}$ is its error generator. The probability distribution over outcomes of the $i$-inverse circuit $\CC^{(i)}$ is, therefore, approximately equal to the probability distribution given by the original circuit $\CC$ plus contributions from error generators corresponding to the pair of extra layers introduced by the local inversion but modified by the unitary operators preceding and following them.
Interestingly, the operator
\begin{equation}\label{eq: Lambda operator}
\Lambda_i = \mathscr{E}_i +\mathcal{U}_i\mathscr{E}_i^{(-1)}\mathcal{U}_i^{\dagger}
\end{equation}
in Eq.~\eqref{eq: Delta_inversion_error} shows that under certain conditions the local inversion technique becomes insensitive to some types of error. For example, $\Lambda_i = 0$ whenever $\mathscr{E}_i=\mathscr{E}_i^{(-1)}$ and $\mathscr{E}_i$ anti-commutes with $\mathcal{U}_i$. An example would be a $Z$-axis coherent error on a Pauli $X$ gate. Similarly, $\Lambda_i=0$ if	$\mathscr{E}_i= - \mathscr{E}_i^{(-1)}$ and $\mathscr{E}_i$ commutes with $\mathcal{U}_i$. This is true for, \eg $X$-rotation gates with over-rotation or under-rotation coherent errors, which are exactly canceled by the inversion process.  As an error generator can be decomposed into elementary error generators~\cite{blume-kohout_taxonomy_2022}, this equation for $\Lambda_i$ tells us that we are sensitive to the rate of errors of certain error generators, and not others. But, as we show in~\cref{sec:technique_refinement}, the insertion and averaging over random Pauli layers recovers the local inversion sensitivity to all types of errors.

We can also write explicit expressions for the TVD in the definition of $\eta^{(i)}$ under these approximations,
\begin{equation}\
\label{eq: TVD_with_first_order_error_approx}
\begin{aligned}
	\eta^{(i)} \approx \frac{1}{2}\sum_k \left\vert \Delta_k^{(i)}\right\vert .	
\end{aligned}
\end{equation}
This is the sense in which the error rates of a given layer are amplified by the layer inversion and show up in the TVD-based comparisons we compute of output distributions. 

Similarly, under these approximations, the $i$-ideal circuit  $\CC^{(i,\text{ideal})}$ probability distribution to first-order in $\delta$ would be
\begin{equation}
\pp(k|\CC^{(i,\text{ideal})})\approx  \pp(k|\CC)-\tilde{\Delta}_k^{(i)}+\mathcal{O}(d^2\delta^2),
\end{equation}
where
\begin{equation}
\tilde{\Delta}_k^{(i)} = \langle\bra{k}( \mathcal{U}_{d}\cdots \mathcal{U}_{i+1}) \mathscr{E}_i  ( \mathcal{U}_{i+1}^{\dagger}\cdots \mathcal{U}_{d}^{\dagger}) \CC_0\ket{\rho_0}\rangle.
\end{equation}
Combining the previous two equations with the definition of $\eta^{(i,\text{ideal})}$ (Eq.\eqref{eq: TVD_with_ideal_layer}) we obtain the following expressions:
\begin{equation}
\label{eq: TVD_ideal_layer_lambda_approc}
	\eta^{(i,\text{ideal})} \approx \frac{1}{2}\sum_k \left\vert \tilde{\Delta}_k^{(i)} \right\vert .	
\end{equation}

Note that if the error maps for the target layer and its inverse are similar, one can expect $\langle\bra{k}( \mathcal{U}_{d}\cdots \mathcal{U}_{i+1}) (\mathscr{E}_i)( \mathcal{U}_{i+1}^{\dagger}\cdots \mathcal{U}_{d}^{\dagger}) \CC_0\ket{\rho_0}\rangle \approx \langle\bra{k}( \mathcal{U}_{d}\cdots \mathcal{U}_{i+1}) (\mathcal{U}_{i}\mathscr{E}_i^{(-1)}\mathcal{U}_{i}^{\dagger})( \mathcal{U}_{i+1}^{\dagger}\cdots \mathcal{U}_{d}^{\dagger}) \CC_0\ket{\rho_0}\rangle$, and hence $\Delta_k^{(i)}\approx 2\tilde{\Delta}_k^{(i)}$, in which case
\begin{equation}
\eta^{(i)} \approx \sum_k \left\vert \tilde{\Delta}_k^{(i)} \right\vert \approx 2\eta^{(i,\text{ideal})}.	
\end{equation}

The two cases where one can prove $\Delta_k^{(i)} = 2\tilde{\Delta}_k^{(i)}$ are where $\mathscr{E}_i = \mathscr{E}_i^{(-1)}$ ($\mathscr{E}_i = -\mathscr{E}_i^{(-1)}$) and $\mathscr{E}_i$ commutes (anti-commutes) with $\mathcal{U}_i$.
Even if this equality does not hold, we expect a significant correlation between the two TVDs in the low-noise regime and/or when the inverse of native gates are similar to the original gates.

\section{Process matrices for simulated error model}
\label{app:error_model_Pauli_transfer_matrices}
We list the process matrices used in our error model for simulations presented in the main text. These are derived from GST experiments run on the IBM Q Ourense, and were also used in Ref.~\onlinecite{cincio_machine_2021}. These process matrices are completely-positive trace-preserving (CPTP) estimated maps of the corresponding operations and are given in the Pauli transfer basis.

\begin{align*}
	I &= \left(\begin{matrix}
		    1.0000  & -0.0000 &   0.0000 &   -0.0000 \\
    0.0042  &  0.9943  & -0.0064  &  0.0178 \\
   -0.0033  &  0.0120  &  0.9962  &  0.0186\\
    0.0029 &  -0.0182  & -0.0167  &  0.9928
	\end{matrix}\right)\\
	\sqrt{X} &= \left(\begin{matrix}
	    1.0000  &  0.0000  &  0.0000  & -0.0000 \\
    0.0007  &  0.9988  & -0.0050  & -0.0055 \\
   -0.0010 &  -0.0060  &  0.0167  & -0.9980 \\
   -0.0017 &   0.0065  &  0.9979  &  0.0176
\end{matrix}\right) \\
\end{align*}

\begin{widetext}
\begin{align*}
\textsc{cnot} = \left(\begin{smallmatrix}	
	1.000 &0 &0 &0 &0 &0 &0 &0 &0 &0 &0 &0 &0 &0 &0 &0 \\ 
0.012 &0.973 &0.016 &0.005 &0.005 &-0.002 &0.012 &-0.004 &-0.002 &0.003 &-0.004 &0.002 &-0.010 &0.008 &0.015 &-0.001 \\ 
0.001 &-0.009 &0.004 &-0.003 &-0.002 &0.000 &-0.023 &0.001 &-0.006 &-0.001 &-0.007 &0.003 &0.005 &-0.019 &0.974 &0.003 \\ 
0.002 &0.006 &0.000 &0.003 &-0.005 &-0.001 &0.002 &-0.021 &-0.010 &0.001 &0.003 &-0.010 &-0.001 &-0.007 &0.004 &0.983 \\ 
0.002 &0.001 &0.012 &-0.008 &0.015 &0.964 &0.017 &0.004 &0.001 &0.020 &-0.018 &0.003 &0.048 &0.020 &-0.002 &-0.004 \\ 
0.002 &-0.001 &0.004 &0.002 &0.980 &0.004 &-0.002 &-0.009 &0.018 &0.001 &-0.005 &0.012 &0.021 &0.042 &0.002 &0.005 \\ 
-0.002 &-0.003 &0.041 &0.002 &-0.009 &0.001 &0.005 &-0.018 &-0.005 &-0.002 &0.003 &0.977 &0.014 &-0.003 &0.000 &0.012 \\ 
-0.003 &-0.006 &-0.002 &0.045 &-0.006 &0.019 &0.015 &0.006 &-0.002 &0.022 &-0.968 &-0.001 &-0.006 &0.001 &-0.008 &0.005 \\ 
0.001 &0.007 &-0.004 &0.001 &0.000 &-0.019 &0.017 &-0.001 &0.011 &0.966 &0.019 &0.003 &0.012 &0.009 &-0.002 &-0.005 \\ 
0.001 &0.008 &0.004 &-0.001 &-0.021 &-0.000 &0.002 &-0.011 &0.981 &0.004 &-0.001 &-0.005 &0.014 &0.004 &0.002 &0.010 \\ 
-0.001 &-0.005 &0.007 &-0.002 &0.005 &0.005 &-0.003 &-0.975 &-0.011 &0.002 &0.007 &-0.020 &-0.003 &-0.002 &0.008 &-0.023 \\ 
-0.002 &-0.012 &0.004 &0.006 &0.003 &-0.021 &0.967 &0.001 &-0.005 &0.017 &0.016 &0.007 &0.003 &0.004 &0.021 &0.004 \\ 
-0.002 &-0.003 &-0.001 &0.001 &-0.021 &-0.035 &-0.008 &-0.001 &-0.010 &-0.006 &0.001 &-0.006 &0.987 &0.002 &0.001 &-0.000 \\ 
-0.008 &0.006 &0.012 &-0.001 &-0.043 &-0.020 &-0.003 &0.003 &-0.010 &-0.009 &0.003 &0.008 &0.011 &0.970 &0.016 &0.007 \\ 
0.005 &-0.018 &0.973 &0.003 &-0.004 &-0.009 &0.002 &0.008 &0.002 &0.005 &-0.001 &-0.039 &-0.004 &-0.007 &0.005 &-0.005 \\ 
0.000 &-0.007 &0.005 &0.982 &0.005 &0.002 &-0.008 &0.003 &0.003 &-0.009 &0.040 &0.002 &0.002 &0.005 &0.001 &0.001 \\
    \end{smallmatrix}\right)
\end{align*}
\end{widetext}

Note that to obtain the corresponding CPTP linear maps in the computational basis we use the following equation:
\cite{greenbaum_introduction_2015}:
\begin{equation}
\Lambda\left(\rho_{0}\right) =\frac{1}{d} \sum_{i, j}^{d^2} \operatorname{Tr}\left(\mathds{P}_{j} \rho_{0}\right) R_{i j} \mathds{P}_{i},
\end{equation}
where $d=2^n$ is the Hilbert space dimension, $\mathds{P}_{j} \in \{I, X, Y, Z\}^{\otimes n}$ are elements of Pauli group, and $R$ is the process matrix in the Pauli transfer basis.

In Table~\ref{tab:error_metrics}, we list error metrics (entanglement infidelity and diamond distance to ideal gate) for these process matrices.

\begin{table}[htb!]
    \centering
    \begin{tabular}{|c||c|c|}
    \hline
        Gate label & Ent. Infidelity & 1/2 diamond distance\\
    \hline
        $I$ & $2.8\cdot10^{-3}$& $1.7\cdot10^{-2}$ \\
    \hline
        $\sqrt{X}$ & $8.8\cdot10^{-4}$ &$1.1\cdot10^{-2}$\\
    \hline
        \textsc{cnot} & $1.9\cdot10^{-2}$& $5.0\cdot10^{-2}$\\
    \hline
    \end{tabular}
    \caption{Error metrics for noisy gates (compared to ideal operations) used in the error model used for simulations.}
    \label{tab:error_metrics}
\end{table}

The error metrics indicate that the $\sqrt{X}$ gate has significant coherent noise. The diamond distance is first-order sensitive to coherent errors, while the entanglement fidelity is only second-order sensitive, and both are first-order sensitive to stochastic errors. Therefore, in the absence of coherent errors, we expect these error metrics to be similar, and the two orders of magnitude difference between them for the $\sqrt{X}$ gate is strong indication of coherent noise. 

Moreover, we can explicitly see how there is noise cancellation when the $\sqrt{X}$ and $\sqrt{X}^{-1} = Z(\pi) \sqrt{X} Z(-\pi)$ process matrices are concatenated by examining the following average gate fidelities:
\begin{align}
	F[ (\sqrt{X})^{\rm ideal}, \sqrt{X}\sqrt{X}^{-1}\sqrt{X}] &= 0.9975 \label{eq:f1} \\
	F[ (\sqrt{X}\sqrt{X}\sqrt{X})^{\rm ideal}, \sqrt{X}\sqrt{X}\sqrt{X}] &= 0.9971, \label{eq:f2}
\end{align}
where $(g)^{\rm ideal}$ denotes the ideal version of a gate $g$, and $F(g, \mathcal{E})$ is the average gate fidelity between $g$ and the channel $\mathcal{E}$, which in term of process matrices in the Pauli transfer basis for each of these operations, $\mathcal{R}_{g/\mathcal{E}}$, is $F(g, \mathcal{E})=\nicefrac{(\tr(\mathcal{R}_g^{\sf T}\mathcal{R}_\mathcal{E})+d)}{d(d+1)}$ \cite{Chow_2012}. If there are no error cancellations the two fidelities above will be the same, since both channels involve the application of three $\sqrt{X}$ gates (recall that the $Z(\theta)$ rotations are ideal). However, the fact that the fidelity in \cref{eq:f1} is greater than the one in \cref{eq:f2} indicates that there was cancellation of errors present in the $\sqrt{X}$ channel when it was concatenated with its inverse.

\section{Experimental procedure}
\label{app:details_experiments}
Here we explicitly detail the steps involved in the experiments with the IBM QIP ibmq\_jakarta:
\begin{enumerate}
\item Using the directed acyclic graph representation of the QAOA or QFT circuit, we split  it into several layers, where, if possible, we group $\sqrt{X}$ and \textsc{cnot} gates in the same layers and group the non-Clifford $Z(\theta)$ gates in separate layers.
\item For a circuit with $d$ layers we create $d+1$ circuits, where the first one is the original circuit without layer inversion and each of the remaining $d$ circuits has a different target layer being locally inverted.
\item We create a list composed of $r$ copies of the $d+1$ circuits such that $r(d+1)\leq 300$, which is the maximum number of circuits per job for ibmq\_jakarta. 
\item When random Pauli gates are included, these are randomly selected for each of the $d$ $i$-inverse circuits. We do not apply random Pauli gates to the target layers with only $Z(\theta)$ gates since these are virtual and essentially perfect~\cite{McKay2017}.
\item Finally, we send 6 consecutive jobs (delay between executed jobs is just a few seconds because we had sole access to the QIP), each with 290 (286) QFT (QAOA) circuits and 1024 shots per circuit. Therefore, for each QFT (QAOA) circuit we have 61,440 (67,584) bit strings that are used to estimate their respective output probability distributions.
\end{enumerate}

We execute 60 (66) copies of each QFT (QAOA) circuit to collect a large number of samples for each circuit's output. This could be directly achieved by increasing the number of shots for each measurement from 1024, but we found that there were hardware limitations to doing this. 
The execution of several copies makes the data more affected by temporal instability of the QIP since the experiment is executed over a longer period of time. However, we can post-process the data to detect for significant temporal instability and remove drift-affected data if they are present. This procedure is detailed in~\cref{app:drift}. We found only a minor impact of drift in the experiments reported here, and for the data presented in the main text, no data removal was necessary.

In order to be consistent in the use of the proposed technique, we apply the local inversion to each of the circuit layers, including those with only $Z(\theta)$ gates. However, since the $Z$-rotations in IBM QIPs are virtual gates~\cite{McKay2017}, the target layers, $\LL_i$, with only $Z(\theta)$ gates and their respective extra layers ($[\LL^{-1}_{i}\LL_{i}]^m$) are automatically combined before the circuit is executed in the QIPs. This is done even if the \verb|optimization_level| in the Qiskit function \texttt{execute} is set to zero to avoid any circuit optimization before its execution. 
Consequently, every time a target layer has only $Z$-rotations, the corresponding circuit that is executed is simply the original circuit without any layer inversion. We then end up comparing the probability distribution of the same circuit output  distribution at different times, one obtained at the beginning of the experiment and the other obtained some time later. Interestingly, this gives us a qualitative way to detect parameter drift in the system between different times of the experiment.

\section{Fisher information based analysis}
\label{app:qfim}
In the main text we used TVD as a distance measure between the original and perturbed output distribution. We also mentioned that other choices are possible. Here we introduce and discuss another choice for the distance measure. 

\begin{figure}[htb!]
\includegraphics[width=0.5\textwidth]{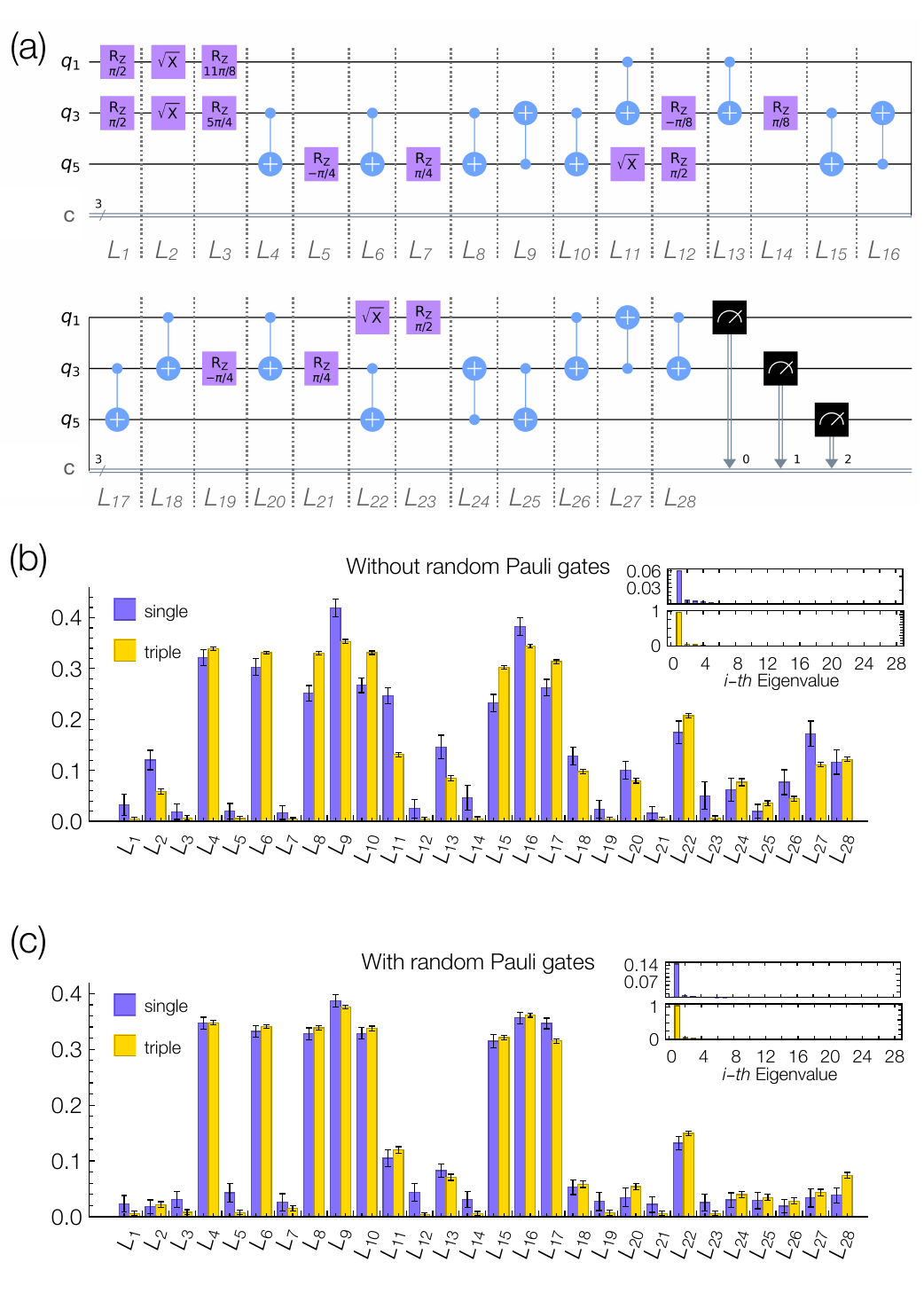}
\caption{\textbf{(a)} 3-qubit QFT circuit transpiled for ibmq\_jakarta, active qubits \{1,3,5\}. \textbf{(b)}  Absolute value of the elements of the eigenvector of the largest qFIM eigenvalue (see inset) for a single layer inversion  (in light purple) and triple layer inversion (in gold). \textbf{(c)} Absolute value of the elements of the eigenvector of the largest qFIM eigenvalue (see inset) for a single layer inversion  (in light purple) and triple layer inversion (in gold) with random Pauli gates. The standard deviation for each element of the eigenvector, indicated with error bars in (b) and (c), is calculated using non-parametric bootstrapping.}
\label{fig:7}
\end{figure}

For a parameterized probability distribution $\pp(x|\vartheta)$ over finite number of outcomes $k$, the Fisher information (FI) is ~\cite{cover_elements_2006}
\begin{equation}
F(\vartheta)=\sum_k \pp(k|\vartheta)\left(\frac{\partial}{\partial \vartheta} \ln \pp(k|\vartheta)\right)^{2},
\end{equation}
which is the variance of $\frac{\partial}{\partial \vartheta} \ln \pp(k|\vartheta)$ (known as the score function). 
Generalizing to the multiparameter case, the Fisher information matrix (FIM) for a probability distribution over sample space $\mathcal{K}$ and parameterized by $m$ parameters, $\pp\left(k|\vartheta_{1}, \ldots \vartheta_{m}\right)$ with $k \in \mathcal{K}$, has elements:
\begin{equation}
\begin{aligned}
F_{i j}\left(\vartheta_{1}, \ldots \vartheta_{m}\right)=&\sum_{k \in \mathcal{K}} \pp\left(k | \vartheta_{1}, \ldots \vartheta_{m}\right)\\
&\times\left[\frac{\partial}{\partial \vartheta_{i}} \log \pp\left(k | \vartheta_{1}, \ldots \vartheta_{m}\right)\right]\\
&\times\left[\frac{\partial}{\partial \vartheta_{j}} \log \pp\left(k | \vartheta_{1}, \ldots \vartheta_{m}\right)\right],
\end{aligned}
\end{equation}
for $1 \leq i, j \leq m$. 

The FI measures the average \emph{sensitivity} of the probability distribution, $\pp$, to variations in its parameters \cite{cover_elements_2006}. In circuit debugging we aim to quantify the sensitivity of the circuit output to a particular layer in the circuit, and therefore the FI seems like it would be a good metric to use for this purpose. However, a quantum circuit's output distribution is not continuous parameterized by its layers, and therefore we introduce a discrete approximation to the FIM, by approximating the derivatives in the definition above with finite differences. We define the quasi-Fisher information matrix (qFIM) with elements:
\begin{equation}\label{eq:elements_quasi_Fisher_information_matrix}
\tilde{F}_{i j}(\CC)=\sum_{k \in B^{n}} \pp(k|\CC)\left[\log \frac{\pp\left(k | \CC^{(i)}\right)}{\pp(k | \CC)}\right]\left[\log \frac{\pp\left(k | \CC^{(j)}\right)}{\pp(k | \CC)}\right].
\end{equation}
Here, $\pp\left(k | \CC^{(i)}\right)$ is the probability distribution that corresponds to the circuit with the $i$-th layer inverted ($\CC^{(i)}$), and $\pp(k | \CC)$ is the output distribution for the original circuit without any inverted layers. For a $d$-layer circuit the qFIM is, therefore, a $d\times d$ real symmetric matrix whose rows and columns correspond to each inverted layer. In the low-error regime, the perturbed probability distributions, $\pp(k|\CC^{(i)})$, are close to $\pp(k|\CC)$ and therefore this approximation of the derivative is expected to be good. However, if the QIP deviates from the low-error regime, then the qFIM does not have a good interpretation. In current generation QIPs, there is no guarantee of low error, especially due to non-Markovian error modes like parameter drift, which is why in the main text we have used the TVD as the measure of distance between distributions --TVD does not have an interpretation in terms of sensitivity analysis, but it is more robust.

Insight into the layers that influence the output distribution the most is attained by examining qFIM's eigenvalues and eigenvectors, where the largest element (in absolute value) of the eigenvector that corresponds to the largest eigenvalue identifies the most dominant layer in the circuit \cite{sarovar_reliability_2017}. As an example, for the 3-qubit QFT circuit in Fig.~\hyperref[fig:7]{\ref*{fig:7}(a)}, we execute experiments using single and triple local inversion ($m=1$ and $m=3$, respectively) for circuits without and with random Pauli gates (Figs.~\hyperref[fig:7]{\ref*{fig:7}(b)} and \hyperref[fig:7]{\ref*{fig:7}(c)},respectively). The qFIM eigenspectra for the two cases (insets in Figs.~\hyperref[fig:7]{\ref*{fig:7}(b)} and \hyperref[fig:7]{\ref*{fig:7}(c)}) show single dominant eigenvalues, and thus all the useful information is mostly found in the components of the eigenvectors of the dominant eigenvalue. The absolute value of the components of the eigenvectors are represented by the bar charts in Fig.~\hyperref[fig:7]{\ref*{fig:7}(b)}, for the case without random Pauli gates, and in Fig.~\hyperref[fig:7]{\ref*{fig:7}(c)} for the case with random Pauli gates. Figure ~\ref{fig:7} also shows that, in contrast to the TVDs, the multiple local inversion has a damping effect on the layers with $Z$-rotations only. This is a consequence of
the noise amplification of each $i$-inverted layer along with the fact that the layers with $Z(\theta)$ gates are not being inverted, and the fact that we are plotting normalized eigenvectors.

Therefore, the qFIM is an alternative metric to the TVD for analyzing the effects of layer inversions for circuit debugging, especially if the QIP under test is guaranteed to be in the low-error regime.

\section{Drift detection}
\label{app:drift}

Ideally the only difference between the output distributions of a circuit and an $i$-inverted circuit will be due to the amplified noise in layer $i$ resulting from the inversion. However, as discussed in the main text, temporal instability of a QIP, or drift, can lead to differences between these distributions, and thus becomes a confounding factor for our circuit debugging technique. In this appendix we demonstrate, with the experimental data reported in the main text, how techniques from Ref. \onlinecite{rudinger_probing_2019} can be used to detect drift within circuit debugging experiments and filter data if needed.

As mentioned in~\cref{app:details_experiments}, for each circuit, we sent six consecutive jobs to be executed in the QIP. This naturally divides our data into 6 sets, each associated with a different time-stamp (or ``context''). Temporal instability between jobs would imply significant variation in the circuit outcome distributions between contexts. Specifically, we want to know if all of the data (for a given circuit) are drawn from the same underlying probability distribution, which would imply that the circuit outcomes are context independent. We address this question using statistical tools introduced in Ref.~\onlinecite{rudinger_probing_2019}, which are concisely presented below.

Considered $Q$ circuits labeled $q=1,2,\ldots,Q$, each with $M$ possible outcomes labeled $m=1,2,\ldots,M$, that are implemented repeatedly in each of $S$ different contexts labeled $s=1,2,\ldots,S$. For each context $s$, the circuit $q$ defines a probability distribution over the $M$ possible measurement outcomes: $\ppbold_{q,s}=(\pp_{q,s,1},\pp_{q,s,2},\ldots,\pp_{q,s,M})$. In the experiments, we run the circuit $q$ $N_{q,s}$ times in each context $s$ and record the total  counts for each measurement $m$, effectively sampling from each of the $\pp_{q,s}$ distributions and producing a full set of count data denoted $x=\{\xxbold_{q,s}=(\xx_{q,s,1},\xx_{q,s,2},\ldots,\xx_{q,s,M})\}$. 	In terms of the data, a circuit $q$ is context independent iff all of the data are drawn from the same underlying probability distribution $\ppbold_{q,0}$. To test for context dependency, a \textit{statistical hypothesis testing} framework can be used with the base assumption \textit{(null hypothesis)} that all $\ppbold_{q,s}=\ppbold_{q,0}$. For general hypothesis testing, the authors in Ref.~\onlinecite{rudinger_probing_2019} found convenient to use the log-likelihood ratio (LLR) statistic, and derived an expression for the LLR based on the maximum likelihood estimate over the null hypothesis, $\hat{\ppbold}_{q,0}=N_q^{-1}(\xx_{q,1},\xx_{q,2},\ldots,\xx_{q,M})$ with $N_q=\sum_s N_{q,s}$ and $\xx_{q,m}=\sum_s\xx_{q,s,m}$, and the maximum likelihood estimate over the full parameter space, $\hat{\pp}_{q,s}=\xxbold_{q,s}/N_{q,s}$, obtaining: 
\begin{equation}\label{eq:LLR_test_for_individual_circuit}
\lambda_q=-2\sum_{m=1}^M\left[\xx_{q,m}\log\left(\frac{\xx_{q,m}}{N_q}\right) - \sum_{s=1}^S\xx_{q,s,m}
\log\left(\frac{\xx_{q,s,m}}{N_{q,s}}\right)\right].
\end{equation}
With this equation, one can calculate the $p$-value ($p$) of $\lambda_q$ approximated by~\cite{rudinger_probing_2019}:
\begin{equation}\label{eq:p-value_for_lamda_q}
p\approx 1-F_k(\lambda_q),
\end{equation}
where $F_k$ is the $\chi_k^2$ cumulative distribution function and $k=(S-1)(M-1)$ is the number of free parameters in the full parameter space of independent $\ppbold_s$. If the $p$-value of $\lambda_q$ is less than a chosen significance threshold level $\alpha\in (0,1)$, then the null hypothesis is rejected, which means detecting context dependence for circuit $q$. This is called \textit{individual circuit test} (ICT) in Ref.~\onlinecite{rudinger_probing_2019}. Now, a popular choice for $\alpha$ is 5\%, corresponding to a 95\% confidence. However, since we have to implement $Q$ independent hypothesis test, we need to adjust the significance of the individual tests to keep  the probability of false detection in all the $Q$ tests to at most $\alpha=5\%$. This can be done by implementing the ICTs with a Hochberg correction~\cite{hochberg_sharper_1988}. This entails~\cite{rudinger_probing_2019}, first, to sort the $Q$ $p$-values from smallest to largest ($p_{(1)}, p_{(2)},\ldots,p_{(Q)})$, second,  to find the largest integer $r$ (denoted $r_{max}$) such that $p_{(r)}\leq\alpha/(Q-r+1)$, and, third, reject the null hypothesis for all circuits with $p$-values smaller than $p_{\text{threshold}}=\alpha/(Q-r_{max}+1)$.

A complementary test statistic, more sensitive than the ICT to context dependence that is distributed uniformly over all circuits, is the aggregate LLR~\cite{rudinger_probing_2019}: $\lambda_{agg}=\sum_{q=1}^Q\lambda_q$. This is the LLR between the null hypothesis of context independence in all circuits and the all-circuit context dependence model.  When the null hypothesis holds, $\lambda_{agg}$ approximately follows a $\chi_{k_{agg}}^2$ distribution with $k_{agg}=Q(S-1)(M-1)$. Since, for $k\gg1$,  the $\chi_k^2$ distribution is approximately normal with mean $k$ and variance $1/(2k)$, it is convenient to use the number of  standard deviations,
\begin{equation}\label{eq:standard_deviation_lambda_agg}
\mathcal{N}_{\sigma}=\frac{\lambda_{agg}-k_{agg}}{\sqrt{2k_{agg}}},
\end{equation}
by which $\lambda_{agg}$ exceeds its expected context-independent value in order to express $\lambda_{agg}$ statistical significance. It is also useful to have a threshold of $\alpha$ significance of the $\mathcal{N}_{\sigma}$:
\begin{equation}\label{eq:standard_deviation_threshold_alpha}
\mathcal{N}_{\sigma,\text{threshold}}=\frac{F_{k_{agg}}^{-1}(1-\alpha)-k_{agg}}{\sqrt{2k_{agg}}},
\end{equation}
where $F_k^{-1}$ is the inverse cumulative distribution function for the $\chi_k^2$ distribution.

Given that the ICTs indicate which circuits vary but are often less sensitive than the aggregate LLR, we can take advantage of these two test statistics by implementing both tests with significance levels adjusted appropriately. To do so, we implement the following \textit{two-step strategy} proposed in Ref.~\onlinecite{rudinger_probing_2019}. For a global significance $\alpha$, $(i)$, implement the aggregate LLR test at a significance level $\alpha/2$. If context dependence is detected, set $\beta=\alpha$; otherwise, set $\beta=\alpha/2$. $(ii)$, implement the ICTs using a Hochberg correction at a significance of $\beta$. If there are more than two contexts in the data ($S_{all}>2$), we can also implement a pairwise comparison between different pairs of contexts, where we implement the above procedure more than once with $S=2$, as well as a joint  comparison of all contexts with $S=S_{all}$. And so, in order to maintain a global significance of $\alpha$, we can split $\alpha$ over each implementation of the above procedure (this is known as \textit{generalized Bonferroni correction}~\cite{lehmann_testing_2005}).

Now, in the main text we presented 8 experiments:
\begin{enumerate}
\item Four 3-qubit QFT experiments.
	\begin{enumerate}
	\item QFT circuits with single local inversion without random Pauli gates.
	\item QFT circuits with triple local inversion without random Pauli gates.
	\item QFT circuits with single local inversion with random Pauli gates.
	\item QFT circuits with triple local inversion with random Pauli gates. 
	\end{enumerate}	
\item Four 4-qubit QAOA experiments.
	\begin{enumerate}
	\item QAOA circuits with the set of optimized angles and triple local inversion, without random Pauli gates.
	\item QAOA circuits with the set of random angles and triple local inversion, without random Pauli gates.
	\item QAOA circuits with the set of optimized angles and triple local inversion, with random Pauli gates.
	\item QAOA circuits with the set of random angles and triple local inversion, with random Pauli gates.
	\end{enumerate}
\end{enumerate}
Each experiment data is divided in 6 \textit{contexts} (data sets). We want to test for drift  in each experiment in two ways: jointly comparing the six contexts and, also, comparing between different pairs of contexts. This results in 13 comparisons between contexts in total. To maintain a global significance of 5\%, therefore, we perform each comparison between contexts at a significance of (5/13)\% $\approx$ 0.38\%. We implement the aforementioned two-step strategy, where, accordingly, the aggregate LLR test is performed at (5/26)\% $\approx$ 0.19\% significance, and the ICT tests are then performed using the Hochberg correction.

\begin{figure}[htb!]
\includegraphics[width=0.5\textwidth]{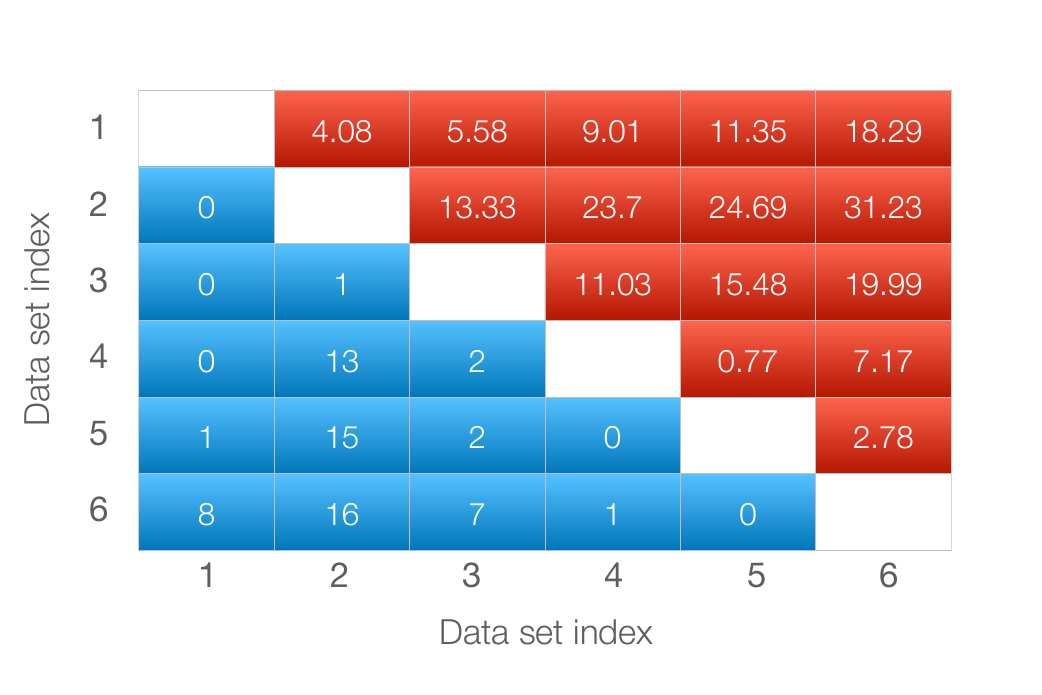}
\caption{An example using the statistical tools for detection of context-dependent errors introduced in Ref.~\onlinecite{rudinger_probing_2019} applied to the experimental results of the QAOA with optimized angles and without random Pauli gates. The data set index correspond to the 6 jobs (contexts) sequentially submitted to ibmq\_jakarta. The upper triangle gives the $\mathcal{N}_{\sigma}$, Eq.~\eqref{eq:standard_deviation_lambda_agg}, for pairwise comparisons between the 6 data sets, the threshold for drift detection for each pair is $\mathcal{N}_{\sigma,\text{threshold}}=3.07$. The lower triangles indicates the number of circuits of the total 26 circuits that are found to contain statistically significant drift between the pair of data sets indicated by the table indexes.}
\label{fig:8}
\end{figure}
\begin{figure}[htb!]
\includegraphics[width=0.5\textwidth]{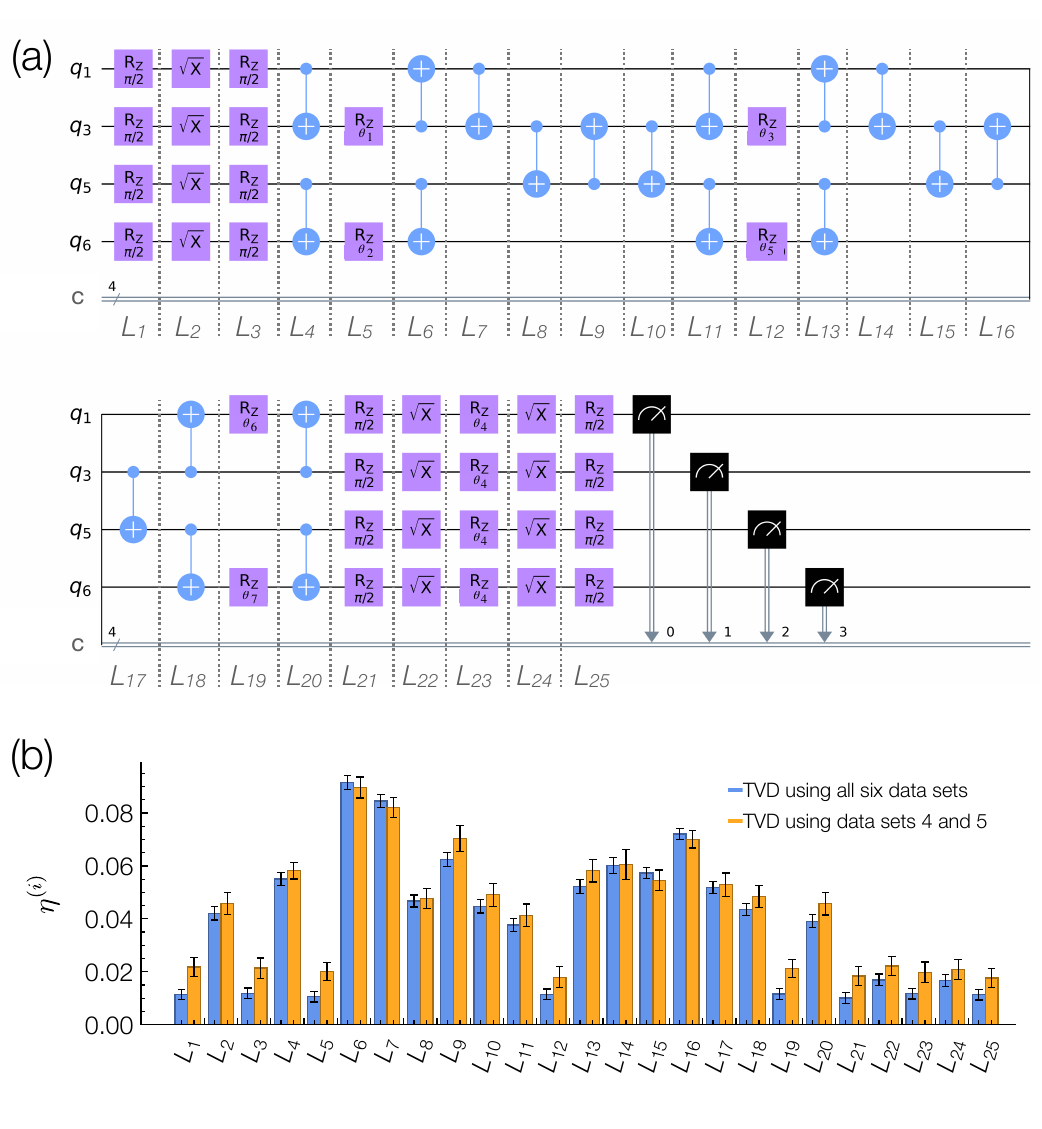}
\caption{\textbf{(a)} 4-qubit QAOA circuit transpiled for ibmq\_jakarta. \textbf{(b)} Comparison of TVDs for the experimental results of the QAOA circuit with optimized angles, using either all six data sets (jobs) or only the data sets 4 and 5, which present no detectable drift between them.}
\label{fig:9}
\end{figure}

For each of the QFT experiments, there are $Q=29$ circuits, each with $M=8$ possible outcomes that are implemented $N_{q,s}=10,240$ in each context.  For the joint comparison of all six contexts, we find for the aggregate LLR test that $\mathcal{N}_{\sigma}$ for the QFT experiments in the above list are: 1.(a)  $\mathcal{N}_{\sigma}\approx 107.23$, 1.(b)  $\mathcal{N}_{\sigma}\approx 270.8$, 1.(c)  $\mathcal{N}_{\sigma}\approx 23.14$, 1.(d)  $\mathcal{N}_{\sigma}\approx 31.59$; the threshold for drift detection is only $\mathcal{N}_{\sigma,\text{threshold}}\approx 3.0$. Therefore, we detect drift in all four QFT experiments with high confidence. The ICTs also detects drift in each QFT experiment, finding 29 circuits in both 1.(a) and 1.(b), 12 circuits in 1.(c), and 23 circuits in 1.(d), to be significant.

Similarly, for each of the QAOA experiments, there are $Q=26$ circuits, each with $M=16$ possible outcomes that are implemented $N_{q,s}=11,264$ in each context. For the joint comparison of all six contexts, we find for the aggregate LLR test that $\mathcal{N}_{\sigma}$ for the QAOA experiments in the above list are: 2.(a)  $\mathcal{N}_{\sigma}\approx 29.62$, 2.(b)  $\mathcal{N}_{\sigma}\approx 30.62$, 2.(c)  $\mathcal{N}_{\sigma}\approx 25.71$, 2.(d)  $\mathcal{N}_{\sigma}\approx 75.19$; the threshold for drift detection is $\mathcal{N}_{\sigma,\text{threshold}}\approx 2.97$. Therefore, we detect drift in all four QAOA experiments with high confidence. The ICTs also detects drift in each QAOA experiment, finding 26 circuits in 2.(a), 25 circuits in 2.(b), 17 circuits in 2.(c), and 26 circuits in 2.(d), to be significant.

The pairwise comparison between contexts for each experiment gives a more detailed picture of the presence of drift in the circuits. Figure~\ref{fig:8} shows an example of the pairwise context comparison results obtained for the QAOA with optimized angles and without random Pauli gates. The upper triangle (in red) of the plot shows the $\mathcal{N}_{\sigma}$ for each pairwise comparison; the threshold for drift detection for each pair is $\mathcal{N}_{\sigma,\text{threshold}}=3.07$. The lower triangle (in blue) of the plot shows the number of circuits that are found to have statistical significant drift for each pairwise comparison. Drift is not detected between a pair of contexts if the number of circuits found to have statistical significant drift is zero and $\mathcal{N}_{\sigma}< \mathcal{N}_{\sigma,\text{threshold}}$; drift is detected otherwise. From the results in Fig.~\ref{fig:8}, we can see that drift is not detected only between the data set pairs $\{4,5\}$ and $\{5,6\}$, the former having the smallest standard deviation. However, when we compare the TVD obtained from averaging all six data sets with the TVD obtained from averaging only the data sets 4 and 5 (with no detectable drift in between), we find a correlation  $>0.99$  between them and, as shown in Fig.~\hyperref[fig:9]{\ref*{fig:9}(b)}, differences larger than the sum of their respective TVD standard deviations are found exclusively in the layers with only $Z$ rotations. The large differences in these \textit{Z-layers}  is not surprising because we already mentioned in~\cref{app:details_experiments} that these layers give us a qualitative way to detect parameter drift in the system between different times of the experiment (contexts) and indeed statistical significant drift has  been formally detected with the tools presented in this appendix. What is more relevant for the debugging purpose of our technique are the layers with no $Z$ rotations. The TVDs of those \textit{non-Z layers} in Fig.~\hyperref[fig:9]{\ref*{fig:9}(b)} do not show differences larger than the sum of their respective standard deviations, and thus the results obtained using all six data sets are not significantly different from the results that we would obtain if we decide to keep only the data sets 4 and 5. In other words, the TVD results obtained using the average of all six data sets are not appreciably affected by drift. Interestingly, we get to the same conclusion when we apply the same statistical analysis to the other seven experiments. Because of this, and the fact that, in general, statistical variation decreases with the size of the data set, we decided to keep the six data sets in the results presented in the main text.

Finally, note that  for each QAOA (QFT) experiment, each job is composed of 11 (10) repetitions per circuit, each with 1024 shots, and thus we can associate each measurement with a different time-stamp or context. In fact, we applied the same statistical analysis to the data sets within each job and found that  statistical significant drift errors were much more scarce than those from the comparison between jobs (therefore, the temporal instability on this device is more significant at timescales longer than a single job), and their impact on the averaged results was also found to be negligible.

\end{appendix}

\end{document}